\documentstyle[12pt]{article}
\reversemarginpar
\newtheorem{lemma}{Lemma}

\newcommand{\nwc}{\newcommand}

\newtheorem{prop}{Proposition}

\nwc{\ba}{\begin{array}}
\nwc{\be}{\begin{equation}}
\nwc{\bea}{\begin{equation}}
\nwc{\beq}{\begin{eqnarray}}
\nwc{\beqn}{\begin{eqnarray*}}
\nwc{\beqast}{\begin{eqnarray*}}
\nwc{\nn}{\nonumber}
\nwc{\bm}{\boldmath}
\nwc{\vep}{\varepsilon}
\nwc{\ep}{\varepsilon}
\nwc{\veps}{\varepsilon}
\nwc{\vrho}{\varrho}
\nwc{\eps}{\epsilon}

\newcommand{\bean}{\begin{eqnarray*}}
\newcommand{\eean}{\end{eqnarray*}}

\newcommand{\commentout}[1]{}

\nwc{\m}{\mbox}
\nwc{\re}{\hbox{Re}}
\nwc{\lamb}{\lambda_\varepsilon}
\nwc{\ls}{\stackrel{<}{\sim}}
\nwc{\gs}{\stackrel{>}{\sim}}
\nwc{\ubm}{\unboldmath}
\nwc{\mt}{\bar{t}}
\nwc{\bmxi}{\m{\bm $\xi$\ubm}}
\nwc{\bmeta}{\m{\bm $\eta$\ubm}}
\nwc{\bma}{\m{\bm $a$\ubm}}
\nwc{\bmb}{\m{\bm $b$\ubm}}
\nwc{\bmc}{\m{\bm $c$\ubm}}
\nwc{\bmd}{\m{\bm $d$\ubm}}
\nwc{\bme}{\m{\bm $e$\ubm}}
\nwc{\bmf}{\m{\bm $f$\ubm}}
\nwc{\bmg}{\m{\bm $g$\ubm}}
\nwc{\bmh}{\m{\bm $h$\ubm}}
\nwc{\bmi}{\m{\bm $i$\ubm}}
\nwc{\bmj}{\m{\bm $j$\ubm}}
\nwc{\bmk}{\m{\bm $k$\ubm}}
\nwc{\bml}{\m{\bm $l$\ubm}}
\nwc{\bmn}{\m{\bm $n$\ubm}}
\nwc{\bmo}{\m{\bm $o$\ubm}}
\nwc{\bmp}{\m{\bm $p$\ubm}}
\nwc{\bmq}{\m{\bm $q$\ubm}}
\nwc{\bmr}{\m{\bm $r$\ubm}}
\nwc{\bms}{\m{\bm $s$\ubm}}
\nwc{\bmt}{\m{\bm $t$\ubm}}
\nwc{\bmu}{\m{\bm $u$\ubm}}
\nwc{\bmv}{\m{\bm $v$\ubm}}
\nwc{\bmw}{\m{\bm $w$\ubm}}
\nwc{\bmx}{\m{\bm $x$\ubm}}
\nwc{\bmxt}{\m{\bm $x$\ubm}^\varepsilon (t)}
\nwc{\bmy}{\m{\bm $y$\ubm}}
\nwc{\bmz}{\m{\bm $z$\ubm}}

\nwc{\bmX}{\m{\bm $X$\ubm}}
\nwc{\bmB}{\m{\bm $B$\ubm}}
\nwc{\bmS}{\m{\bm $S$\ubm}}
\nwc{\bmR}{\m{\bm $R$\ubm}}
\nwc{\bmU}{\m{\bm $U$\ubm}}
\nwc{\bmE}{\m{\bm $E$\ubm}}
\nwc{\bmF}{\m{\bm $F$\ubm}}
\nwc{\bmH}{\m{\bm $H$\ubm}}
\nwc{\bmI}{\m{\bm $I$\ubm}}
\nwc{\bmP}{\m{\bm $P$\ubm}}
\nwc{\bmM}{\m{\bm $M$\ubm}}
\nwc{\bmJ}{\m{\bm $J$\ubm}}
\nwc{\bmK}{\m{\bm $K$\ubm}}
\nwc{\bmA}{\m{\bm $A$\ubm}}
\nwc{\bmD}{\m{\bm $D$\ubm}}
\nwc{\bmG}{\m{\bm $G$\ubm}}

\nwc{\bmtheta}{\m{\bm $\theta$\ubm}}
\nwc{\bmnu}{\m{\bm $\nu$\ubm}}
\nwc{\bmom}{\m{\bm $\omega$\ubm}}
\nwc{\om}{\omega}
\nwc{\Om}{\m{\bm $\Omega$\ubm}}
\nwc{\bmsigma}{\m{\bm $\sigma$\ubm}}
\nwc{\bmnabla}{\m{\bm $\nabla$\ubm}}
\nwc{\bmLambda}{\m{\bm $\Lambda$\ubm}}
\nwc{\bmlambda}{\m{\bm $\lambda$\ubm}}
\nwc{\bmtau}{\m{\bm $\tau$\ubm}}
\nwc{\bmPhi}{\m{\bm $\Phi$\ubm}}
\nwc{\bmphi}{\m{\bm $\phi$\ubm}}
\nwc{\bmPsi}{\m{\bm $\Psi$\ubm}}
\nwc{\bmpsi}{\m{\bm $\psi$\ubm}}
\nwc{\bmGamma}{\m{\bm $\Gamma$\ubm}}
\nwc{\bGamma}{\m{\bm $\Gamma$\ubm}}
\nwc{\bmgamma}{\m{\bm $\gamma$\ubm}}
\nwc{\bmQ}{\m{\bm $Q$\ubm}}

\nwc{\bE}{{\bf E}}
\nwc{\bW}{{\bf W}}
\nwc{\bF}{{\bf F}}
\nwc{\bD}{{\bf D}}
\nwc{\bJ}{{\bf J}}
\nwc{\bK}{{\bf K}}
\nwc{\bI}{{\bf I}}
\nwc{\bG}{{\bf G}}
\nwc{\bA}{{\bf A}}
\nwc{\bZ}{{\bf Z}}
\nwc{\bB}{{\bf B}}
\nwc{\bH}{{\bf H}}
\nwc{\bR}{{\bf R}}
\nwc{\ia}{{\it a}}
\nwc{\bC}{{\bf C}}
\nwc{\bx}{{\bf x}}
\nwc{\bfe}{{\bf e}}
\nwc{\by}{{\bf y}}
\nwc{\bk}{{\bf k}}
\nwc{\bp}{{\bf p}}
\nwc{\bS}{{\bf S}}
\nwc{\bi}{{\bf i}}
\nwc{\bw}{{\bf w}}

\newcommand{\vV}{{\bf V}}
\newcommand{\vU}{{\bf U}}

\nwc{\cA}{{\cal A}}
\nwc{\cV}{{\cal V}}
\nwc{\cL}{{\cal L}}
\nwc{\cB}{{\cal B}}
\nwc{\cC}{{\cal C}}
\nwc{\cD}{{\cal D}}
\nwc{\cO}{{\cal O}}
\nwc{\cao}{{\cal A}^{-1}}
\nwc{\cE}{{\cal E}}
\nwc{\cf}{{\cal F}}
\nwc{\cg}{{\cal G}}
\nwc{\cH}{{\cal H}}
\nwc{\cQ}{{\cal Q}}
\nwc{\cI}{{\cal I}}
\nwc{\cJ}{{\cal J}}
\nwc{\cK}{{\cal K}}
\nwc{\cl}{{\cal L}}
\nwc{\cle}{{\cal L}^\varepsilon}
\nwc{\clu}{{\cal L}{\cal U}}
\nwc{\cm}{{\cal M}}
\nwc{\cn}{{\cal N}}
\nwc{\co}{{\cal O}}
\nwc{\cp}{{\cal P}}
\nwc{\cpt}{{\cal P}^\varepsilon_t}
\nwc{\cq}{{\cal Q}}
\nwc{\calr}{{\cal R}}
\nwc{\cs}{{\cal S}}
\nwc{\ct}{{\cal T}}
\nwc{\cu}{{\cal U}}
\nwc{\cv}{{\cal V}}
\nwc{\cw}{{\cal W}}
\nwc{\cx}{{\cal X}}
\nwc{\cy}{{\cal Y}}
\nwc{\cz}{{\cal Z}}

\nwc{\ea}{\end{array}}
\nwc{\ee}{\end{equation}}
\nwc{\eea}{\end{equation}}
\nwc{\eeq}{\end{eqnarray}}
\nwc{\eeqn}{\end{eqnarray*}}
\nwc{\eeqast}{\end{eqnarray*}}

\nwc{\noi}{\noindent}
\nwc{\non}{\nonumber}
\nwc{\half}{\frac{1}{2}}
\nwc{\third}{\frac{1}{3}}

\nwc{\uP}{{\em \bf Proof: }}
\nwc{\uT}{\underline{Theorem:}}

\begin{document}
\setcounter{page}{1}
\title{
Phase Diagram for Turbulent Transport:
Sampling Drift, Eddy Diffusivity and Variational Principles}

\author{Albert C. Fannjiang\\
\small
Department of Mathematics, 
University of California at Davis\\
}

\maketitle

{\bf Abstract.} 
We study the long-time, large scale
transport in a three-parameter family
of isotropic, incompressible velocity fields with power-law spectra.
Scaling law for transport is characterized by 
the scaling exponent $q$ and the
Hurst exponent $H$, as functions of the parameters. 
The parameter space is divided into regimes of scaling laws of different
{\em functional forms} of the scaling exponent and the Hurst exponent.
We present
the full three-dimensional phase diagram.

The limiting process is one of three kinds:
Brownian motion ($H=1/2$), 
persistent
fractional Brownian motions ($1/2<H<1$) and regular (or smooth) motion ($H=1$).

We discover that a critical wave number divides the infrared cutoffs
into three categories, critical, subcritical and supercritical; they
give rise to different scaling laws and phase diagrams.
We introduce the notions of 
sampling drift and eddy diffusivity, and formulate variational principles
to estimate the eddy diffusivity. We show that fractional Brownian motions
result from a dominant sampling drift. 

\bigskip

{\bf 1991 MSC code.} Primary 76M35, 76R50; Secondary 76F05, 60G60.

\bigskip

{\bf Keywords.}
Sampling drift, critical wave number, eddy diffusivity,  variational principles,
fractional Brownian motion.

\newpage
\tableofcontents
\newpage
\section{Introduction}
The movement of a passive scalar in a turbulent flow is described
by the stochastic differential equation
\[
d\bmx(t)=\vV(\bmx(t),t, \omega)dt+\sqrt{2\kappa}d\bmw(t), \quad
\bmx(0)=0,
\]
where $\bmx(t)$ is the position of the particle at time $t$, $\kappa\geq 0$ the molecular
diffusivity, $\bmw(t)$ the standard
Brownian motion and $\vV(\bx,t, \omega)$ a time-stationary, 
space-homogeneous, incompressible velocity field. Here $\omega$ denotes
an element of an ensemble of random flows.

We are concerned with the long time, large scale behavior
of the displacement $\bmx(t)$. To this end,  we study the scaling limit 
\be
\label{2.1}
\bmx^\vep(t)=\vep\bmx(t/\vep^{2q}), \quad\hbox{as $\ep\to 0$},
\ee
with suitable $q>0$. The scaling exponent $q$ characterizes
the time scale associated with transport on the spatial observation
scale $1/\veps$. The equation  for the rescaled displacement (\ref{2.1})
becomes
\be
\label{2.2'}
\nonumber
d\bmx^\vep(t)=\vep^{1-2q}\vV(\bmx^\vep(t)/\vep,
t/\vep^{2q}) dt+
\vep^{1-q}\sqrt{2\kappa}d\bmw(t).
\ee
When molecular diffusion is evidently negligible, we set
$\kappa=0$ to simplify the equation
\be
\label{2.2}
d\bmx^\ep(t)=
\ep^{1-2q}\vV(\bmx^\ep(t)/\ep,t/\ep^{2q})dt.
\ee
The effect of molecular diffusion is discussed where the issue
arises and in the
concluding remarks.

Motivated by existing diffusion limit theorems
for steady flows with finite-range, spatial correlations 
(\cite{dt},\cite{moser2}),
on one hand,
and those for temporally mixing flows with long-range, spatial correlations
(\cite{CX}, \cite{markov}, \cite{Kor}), on the other hand, we
consider turbulent transport in a class of random flows
with power-law spectra parametrized by $\alpha,\beta,\gamma$
(see Section 2 for details).

Roughly speaking, the velocity field $\vV$ is 
time-stationary, space-homogeneous and Gaussian. Its
two-point correlation functions $\bR=[R_{ij}],
R_{ij}(\bx,t)=\langle U_i(\cdot,\cdot)U_j(\cdot+\bx,\cdot+t)\rangle$,
are given by the Fourier transform
$R_{ij}(\bx,t)=\int e^{i\bk\cdot\bx}\widehat{R}_{ij}(\bk,t)d\bk$ with
\be
\label{1.1}
\widehat{R}_{ij}(\bk,t)=
\rho(|\bk|^{2\beta}t)\cE(\bk)\left(\delta _{ij}-
k_{i}k_{j}|\bk|^{-2}\right)|\bk|^{1-d} 
\ee
where $\rho$ is the time correlation (relaxation) function and 
$\cE$ the energy (shell) spectrum given  by
a power-law
\[
\cE(\bk)=
E_0|\bk|^{1-2\alpha},
\quad E_0>0.
\]
Here $\langle\cdot\rangle$ denotes the ensemble average.
The factor $\left( \delta _{ij}-
k_{i}k_{j}|\bk|^{-2}\right) 
$ ensures that the flow is divergence-free.
If $\rho$ is an {\em exponential} function, $e^{-a_0|\bk|^{2\beta}t}$, then
the velocity field is an Ornstein-Uhlenbeck process which is Markovian.

Note that the spectrum is not integrable near $|\bk|=
\infty$ or
$|\bk|=0$ for 
$\alpha \leq 1$ or $
\alpha\geq 1$, respectively. 
The infrared divergence (small $|\bk|$) of the integral of velocity
 energy spectrum indicates non-homogeneous velocity and thus
violates the space homogeneity assumption, whereas the ultraviolet divergence
(large $|\bk|$) of the integral would make the velocity  a generalized,
rather than ordinary, 
function (i.e. a distribution).
To remove divergence in the spectrum,
we introduce 
an ultraviolet cutoff 
\be 
\label{1.2}
\nonumber
{\cal E}(\bk)=0,
\quad |\bk| >K,
\quad\hbox{for  }
\alpha\leq 1
\ee
and an infrared cutoff 
\be
\label{1.3}
{\cal E}(\bk)=0,
\quad |\bk|<\delta\ll 1,
\quad\hbox{for  } \alpha \geq 1.
\ee
In the case of $\alpha <1$, the energy containing scale is at the
ultraviolet cutoff; in the case of $\alpha>1$, 
the energy containing scale is at the infrared cutoff.
It is convenient to write the cut-off energy spectrum 
as
\be
\label{2.41}
{\cal E}(\bk)=E_0|\bk|^{1-2\alpha}I(|\bk|)
\ee
where $I(|\bk|)$ is the characteristic function of
$
[0,K]$,
for $\alpha<1$, of $[\delta,\infty)$ for $\alpha> 1$, and
of $[\delta,K]$, for $\alpha=1$ (see \cite{Fr}).
When we study the effect of an infrared cutoff, 
we will take an infrared cutoff $\delta=\ep^\gamma>0$.
Ultraviolet cutoffs play less prominent
role than do infrared cutoffs in the scaling limit.

If the infrared cutoff $\delta>0$ is fixed, independent of $\epsilon$,
then the
flow is mixing in time (i.e. correlation time is uniformly bounded,
independent of
wave number) and, consequently, 
the scaling in (\ref{2.1}) is diffusive, $q= 1$,  and the limit is
a Brownian motion (\cite{markov}).
In the case of the Kolmogorov-Obukhov spectrum
($\alpha=4/3,\beta=1/3$, see \cite{Fr}), $[\delta, K]$ represents the
inertial range where $K^{-1}$ is the dissipation length and $\delta^{-1}$
is the integral length. In general,
the infrared cutoff is determined by the scale of
external forcing and the size of physical
domain. By letting $\delta$ change with $\epsilon$, as $\delta=
\ep^\gamma, \gamma>0$, we vary
the spatial scale of observation $1/\epsilon$ in relation to, e.g., 
the size of
physical domain.

If the scaling limit exists, {\em statistically} independent of the
initial point, and has
stationary increments,
then the transport process is said to be {\em homogenized},
and the up-scaling, or coarse-graining, procedure represented by (\ref{2.1})
is called {\em homogenization}.
The scaling is 
diffusive if $q=1$, superdiffusive if $q<1$, subdiffusive
if $q>1$. Sub- and super-diffusions are called anomalous diffusion.

The limit $\bZ(t)$ may be Gaussian or non-Gaussian,
Markovian or non-Markovian, even if the velocity field is Gaussian
and Markovian.
In general $\bZ(t)$
has stationary increments as does $\bmx^\ep(t)$ (\cite{PS}).
If $\bZ(t)$ is {\em self-similar} and 
Gaussian then it can be characterized by a unique
Hurst exponent $H$ in its autocovariance function
\be
\label{H'}
\hbox{Cov}(\bZ(t_1),\bZ(t_2))=
{1\over 2}\bC\left\{|t_1|^{2H}+|t_2|^{2H}-|t_1-t_2|^{2H}\right\},\quad
\quad 0 < H\leq 1
\ee
where $\bC$ is the variance of $\bZ(1)$.
$H=1$ corresponds to a {\em regular} (or {\em smooth}) motion;
$H=1/2$ a Brownian motion, $\bB(t)$.
Any other $H$ corresponds  to a
fractional Brownian motion (FBM), $\bB_H(t)$, 
which, after normalization, can be represented as
\be
\label{fbm}
\bB_H(t)=
\int^0_{-\infty}(|t-t'|^{H-1/2}-|t'|^{H-1/2})d\bB(t')
+\int^t_0|t-t'|^{H-1/2}d\bB(t'),\quad 0<H<1,
\ee
as introduced in \cite{MV}.
Eq. (\ref{fbm}) defines the only mean-zero, mean-square, continuous,
Gaussian process that is self-similar (or self-affine),
with the Hurst exponent $H$, has stationary increments, and
satisfies $\bB_H(0)=0$ (see \cite{ST}). FBMs found in the present
study are all persistent in the sense that $H>1/2$.
It is worth noting that, for $\alpha\geq 1$ with the critical cutoff
($\gamma=\gamma_c$), the limit process
is not self-similar (cf. Regimes II', III' and IV').

Non-Markovian limits are related to non-local
homogenization (\cite{BLP}, \cite{E}, \cite{T1}, \cite{T2}). Previously
nonlocal homogenization has been  shown
to arise as a result of fast oscillation,
rather than of a scaling limit.

If $\bZ(t)$ is not Gaussian, then there may be a hierarchy of Hurst exponents
corresponding to higher moments of the process. When the sequence of
Hurst exponents diverges as the order of moment increases, the limit
is {\em intermittent}.
Intermittency effect
may also manifest as multiple scaling exponents.
We do not consider the problem of intermittency here. 

In this paper,
we do not address directly the question of existence and uniqueness
of the scaling limit. Rather, we assume the
existence and uniqueness of a nontrivial scaling limit,
and
seek to identify the scaling exponent
and the second order Hurst exponent (the Hurst exponent,
in case of a Gaussian limit). In doing so, we point out
relevant mathematical results that exist, or can be or are yet to
be proved.
We try to present a simple, coherent physical picture of
the whole phase diagrams. To enhance our case,
we often analyze the problems from several different perspectives.

The exponent $q$ characterizes the time scale associated with transport
observed on the space scale $1/\ep$; the exponent $H$ characterizes
the time correlation property of successive increments on the observation
scale
(and, therefore, the roughness of the limiting sample paths). 
Naturally we ask if the dimensionally correct relation
\be
H=1/(2q)
\label{H}
\ee
holds?
When (\ref{H}) holds the
limit process is invariant under the same scaling transformation (\ref{2.1}).
It turns out that relation (\ref{H}) generally holds for $\alpha<1$ but fails
for $\alpha>1$.
If an additional infrared cutoff is made in the case of $\alpha<1$
and if the cutoff is removed faster
than some {\em critical} wave number, $k_c$,
then (\ref{H}) does not hold. In these situations, the inequality
\be
\label{H2}
H<1/(2q)
\ee
is in place of (\ref{H}). The inequality (\ref{H2}) is due to the fact that
$2H$ characterizes the {\em  covariances}, 
whereas $1/q$ characterizes the {\em variances},
of successive
increments of turbulent motion on the observation scale.

In general the exponents $q, H$ depend on the parameters $\alpha, \beta$ and
the cutoff $\delta=\ep^\gamma$ and can be expressed explicitly as functions of
$\alpha,\beta,\gamma$. 
Here it may be helpful to draw analogy to 
critical phenomena in statistical physics: we think of
$\alpha,\beta,\gamma$ are order parameters and the scaling limit
$\ep\to 0$ as thermodynamic limit and the exponents
$q,H$ given by formulas of $\alpha,\beta,\gamma$
as phases. The phase diagram divides
the space of order parameters $\alpha,
\beta, \gamma$ into regions associated with
different formulas for $q, H$.
Our results are summarized in Figures 1, 2 , 3 and 4.
Since there are three parameters,
the full phase diagram is three dimensional. To simplify the
presentation, we choose to
portray the full diagram as two two-dimensional diagrams, one for
supercritical and the other for subcritical infrared cutoffs.

Note also that the phase diagrams
are different from those in statistical mechanics in that
our phases are continuum, not discrete:
except for the diffusive regime of $\alpha+\beta<1$ where $H=1/2, q=1$,
$H, q$ change from point to point, continuously or discontinuously. But
their functional forms in relation to
$\alpha,\beta,\gamma$ are discrete and divided by phase boundaries.

Phase diagram was first used by
Avellaneda and Majda \cite{AM0}, \cite{AM2} to present scaling limits
of turbulent transport
in anisotropic, stratified flows of the form $\vV(\bx,t)=(v(x_2,t),0)$,
with $\bx=(x_1,x_2)$. A different diagram 
for the same shear-layer flows was rigorously obtained
by Zhang and Glimm \cite{GZ} using a different approach.
In the current paper we consider 
{\em isotropic}
turbulent flows and the results are different
from those for {\em anisotropic} flows.
Also, we do not attempt to derive the results rigorously here.
Often we refer to existing theorems to indicate
how in principle results may be proved, subject to technical modification,
and to support the physical
arguments invoked; they are not intended to be mathematical proofs.
The proofs of many of the results are very technical
and will be published
elsewhere.

The effect
of an infrared cutoff
depends on whether the cutoff is 
{\em subcritical} 
or {\em supercritical}.
For $\alpha<1$,
a supercritical cutoff, $\gamma>\gamma_c=\max{\{1, 1/(\alpha+2\beta-1)\}}$,
does not affect
the scaling law.
For $\alpha\geq 1$, because the infrared cutoff corresponds to the energy-containing
scale,
the scaling limit is dominated by 
the infrared cutoff.

The supercritical diagram includes:
\begin{itemize}
\item
Regime I: $\alpha+\beta< 1$ or $\alpha<0$.
The scaling is diffusive, $q=1$, and the limit is a Brownian motion,
$H=1/2$.

\item
Regime II: $\alpha+\beta>1,\alpha+2\beta<2, \alpha<1,
\gamma>1/(\alpha+2\beta-1)$. A FBM regime with the space-freezing property
that velocity dependence on space is negligible.
The scaling is superdiffusive,
$q=\beta/(\alpha+2\beta
-1)$, and the limit is a fractional Brownian motion, $H=1/(2q)$.

\item
Regime III: $\alpha+2\beta\geq 2, 0\leq\alpha<1,\gamma>1$.
A FBM regime with the time-freezing property
that velocity dependence on time is negligible.
The scaling is superdiffusive, $q=1-\alpha/2$, and the limit is a fractional 
Brownian motion, $H=1/(2q)$.

\item Regime IV: $1\leq \alpha<2,\gamma>\max{\{1, 1/(\alpha+2\beta-1)\}}$.
A regular (or smooth) motion regime with both the
space-freezing and the time-freezing properties.
The scaling is superdiffusive, $q=(1+\gamma)/2-\gamma\alpha/2$,
and the limit is regular ($H=1$).
\end{itemize}
The relation
(\ref{H}) is satisfied in all but Regime IV.

In the case of 
subcritical cutoffs, $\gamma<\gamma_c=\max{\{1, (\alpha+2\beta-1)^{-1}\}}$,
the number
of regimes shrinks as the significance of low wave numbers
is reduced: part of Regime IV merges with Regime II, and part of
Regime IV merges with Regime III. 
The scaling exponent now depends on
the cutoff exponent $\gamma$ explicitly.
The limit is universally a Brownian motion
across all regimes.

The subcritical regime includes:
\begin{itemize}
\item
Regime I remains intact.

\item Regime V: $\alpha+\beta>1,\alpha+2\beta<2,\gamma<1/(\alpha+2\beta-1)$.
Velocity decorrelation in time dominates the transport.
The scaling is superdiffusive, $q=1+\gamma-\gamma(\alpha+\beta)$.

\item Regime VI: $\alpha+2\beta\geq 2, \alpha\geq 0,\gamma<1$.
Velocity decorrelation in space dominates the transport.
The scaling is superdiffusive, $q=1-\gamma\alpha/2$.
\end{itemize}

Finally, there are three regimes associated with critical cutoffs
for which the limit process is {\em not} self-similar and, thus,
the Hurst exponent is not well-defined.
\begin{itemize}
\item Regime II': $\alpha+\beta>1, \alpha+2\beta<2, 0<\alpha<1$ with
$\gamma=(\alpha+2\beta-1)^{-1}$.
\item Regime III': $\alpha+2\beta>2, 0<\alpha<1$ with $\gamma=1$.
\item Regime IV': $1<\alpha<1+1/\gamma$
with $\gamma=\max{\{1, (\alpha+\beta-1)^{-1}\}}$. 
\end{itemize}

Part of Regimes V and VI
was first
studied by Avellaneda and Majda \cite{AM} (see also
\cite{GS2}). 
The main difference in assumption and setup between this work and \cite{AM}
is that they considered a partial diagram ($0<\beta<1/2, 0<\alpha<2$)
with an infrared cutoff $\gamma=1\leq \gamma_c$
(see also \cite{P}). Figure 4 is a generalization of
theirs.
The phase diagram of \cite{Ka} was obtained
entirely by certain scaling arguments, and is restricted to two dimensions.

By contrast, our main findings are: (i) the transport effect of 
the sampling drift and related critical wave number,
which are introduced for the first time,
(ii) fractional Brownian motion limit as a result of
the critical wave numbers,
(iii) the effect of
infrared cutoffs,
(iv) the formulation of cutoff dependent eddy diffusivity
and its associated variational principle without molecular
diffusion.  The variational principle gives
a useful bound for the 
eddy diffusivity. 

The organization of the paper is as follows. In Section~2, we define the
three-parameter family of Gaussian flows, whose transport properties
are discussed in subsequent
sections.  In Section~3, we introduce the notions of sampling drift,
critical wave numbers 
and eddy diffusivity. We also formulate variational principles
that lead to general bounds for the cutoff dependent eddy diffusivity
in terms of a fractional vector potential of the velocity field.
Since the transition from ultraviolet to infrared cutoff
in velocity occurs at $\alpha=1$, we divide the discussion accordingly into
two cases: $\alpha<1$ and $\alpha>1$. We consider  the case $\alpha<1$ 
in Section~4 and the case $\alpha>1$ in Section~5.  We conclude
with various remarks in Section~6. In Appendix we 
derive a variational principle for the cutoff dependent eddy diffusivity,
without the presence of molecular diffusion.

\section{Random velocity field}
In this section, we describe some mathematical properties of
the random velocity fields considered in this paper.

The most important property is stationarity in time and homogeneity in space
(space-time stationarity for short),
without which homogenization is unlikely to hold.
It should be noted that, when formulated in a general, abstract framework
as we will do momentarily, space-time stationarity encompasses
space-time periodicity, quasi-periodicity and almost periodicity
as well as random stationarity. This abstract formulation is also
handy for formulating the variational principle for the eddy diffusivity
(Section~3.2). Elsewhere, the paper can be understood without
referring to the abstract formulation.

The variational principle in the absence of molecular diffusivity 
also uses explicitly the Markov property of the flow and the
associated generator. A key turbulent diffusion theorem cited
in the discussion of the diffusive regime (Section~4.1) was proved
for certain Markovian velocity fields. For Markovian flows,
the mixing property conveniently corresponds to the spectral gap of the
generator. Elsewhere,
the Markov property is not used explicitly  and probably not needed.

Since we only use the (energy) spectral density explicitly in presenting
the phase diagrams, it is safe to assume that the velocity fields
are Gaussian. In particular,
the Gaussian property is essential
in the fractional-Brownian-motion regimes (II
and III). Elsewhere, the Gaussian property is probably not important.

Let us begin with the abstract formulation of space-time stationarity,
upon which we will define the Gaussian and Markov properties.
Let $\Om$ be the space of {\em steady}, space-homogeneous velocity fields
and let $P$ be a probability measure on
$\Om$.
Homogeneity in space can
be canonically described by the invariance of the distribution $P$ under
the group of translations $\{\bmtau_{\bx}\}_{\bx\in R^d}$ acting on $\Om$.
We further assume that $P$ is ergodic with respect to $\{\bmtau_{\bx}\}_{\bx\in
R^d}$
in the sense that the only invariant, measurable functions on $\Om$ under
$\{\bmtau_{\bx}\}_{\bx\in R^d}$ are constants.
The measure $P$ dictates
the correlation of the velocity field in {\em space},
and, in case of Gaussian velocity fields, is determined by
the energy spectrum.

Alternatively,
we think of $\Om$ as the ensemble of elements $\om$, representing
the randomness of the velocity field, which is distributed
according to the measure $P$.
A (prototypical) random velocity field is a vector-valued, 
random variable (i.e., a function
on $\Om$), denoted by $\widetilde{\vV}(\om)$.
The realization or the sample of the (time independent) velocity
field, $\vV(\bx,\om)$, is the translate of 
$\widetilde{\vV}(\om)$ on $\Om$,
i.e., $\vV(\bx,\om)=\widetilde{\vV}(\bmtau_{\bx}\om)$.
Since the measure $P$ is invariant under the translations,
the resulting velocity fields are space-homogeneous.
We assume that $\widetilde{\vV}$ has zero mean
 \[
 \langle \widetilde{\vV}\rangle =0
 \]
 and zero divergence 
 \[
 \nabla\cdot \widetilde{\vV}=0, \quad\nabla=(\partial_1,\partial_2,...,\partial_d).
 \]
Partial derivative $\partial_i $ is the infinitesimal generator
of  the subgroup of translation $\{\bmtau_{x_i }\}_{x_i \in R}$.
The Laplacian $\Delta:= \nabla\cdot\nabla$
is defined as usual.
As before, $\langle\cdot\rangle$ denotes the ensemble average with
respect to the distribution $P$.

The time dependence of the velocity field
is then introduced as
a time-stationary stochastic process, $\om(t)$, on the space $\Om$,
which preserves the measure $P$. In other words, $P$ is
an invariant measure of the process $\om(t)$.
The realization of time dependent velocity field is then given by
\[
\vV(\bx,t,\om)=\widetilde{\vV}(\bmtau_{\bx}\om(t)),\quad \om(0)=\om.
\]

In this 
formulation, the temporal properties are conveniently separated from
the spatial properties of the velocity field. Additional 
structures such as Gaussianity and Markovianity can be added on
by imposing corresponding properties on $P$ and $\om(t)$.
The space $\Om$ is usually
infinite dimensional in suitable coordinates such as Fourier modes.
This formulation is sufficiently general to describe periodic,
quasi-periodic, almost periodic as well as random homogeneous velocity
fields (see, e.g., \cite{D}).

We think of a Markovian velocity field as a sample path in $\Om$ of
a Markov process $\om(t)$.
A Markovian, Gaussian velocity field corresponds to
an {\em exponential} time correlation function $\rho$ in (\ref{1.1})
and admits the spectral representation
\[
\vV(\bx,t)=\int\limits_{R^d}e^{i2\pi\bk\cdot\bx}\widehat{\vV}(d\bk,t)
\]
where
the stochastic measure $\widehat{\vV}(d\bk,t)$
is an Ornstein-Uhlenbeck process 
\be
\label{12.30}
d_t\widehat{\vV}(d\bk,t)=-a_0|\bk|^{2\beta}
\widehat{\vV}(d\bk,t)dt+|\bk|^\beta
{\cal E}^{1/2}(\bk)
\left(\bI-\bk\otimes\bk|\bk|^{-2}\right)^{1/2}|\bk|^{(1-d)/2}\bW(
d\bk,dt)
\ee
and can be conveniently
expressed in terms of a Gaussian white noise $\bW(d\bk,ds)$
\[
\widehat{\vV}(d\bk,t)=\int\limits_{-\infty}^{t}
e^{-a_0|\bk|^{2\beta}(t-s)}|\bk|^\beta{\cal E}^{1/2}(\bk)
\left(\bI-\bk\otimes\bk|\bk|^{-2}\right)^{1/2}|\bk|^{(1-d)/2}\bW(d\bk,ds).
\]

The Ornstein-Uhlenbeck process
(\ref{12.30}) has an invariant measure $P$ that is a Gaussian distribution
with zero mean and  the variance matrix 
$\bR=[\widehat{R}_{ij}]$ given by (\ref{1.1}).
Then the exponential relaxation function corresponds to
a generator $\cA$ of the form
\be
\label{2.30}
\cA=(-{1\over 4\pi^2}\Delta)^{\beta}\cA_0,\quad \beta\geq 0
\ee
where 
$\cA_0$ is the generator of the process
\beq
\label{5.1'}
d_t\widehat{\vV}_0(d\bk,t)=-a_0\widehat{\vV}_0(d\bk,t)dt+
{\cal E}^{1/2}(\bk)
\left(\bI-\bk\otimes\bk|\bk|^{-2}\right)^{1/2}|\bk|^{(1-d)/2}\bW(
d\bk,dt).
\eeq
The operator $\cA_0$ is symmetric with respect to the measure $P$
and commutes with the translation $\tau_{\bx},\forall \bx\in R^d$.
As the process (\ref{5.1'}) is a time change of
(\ref{12.30}) and different wave numbers are independent,
the measure $P$ remains invariant 
with respect to (\ref{5.1'}).
Also, because the time correlation function for
(\ref{5.1'}) is exponential with an exponent
$a_0$ uniformly bounded above zero, $\cA_0$ has
a spectral gap
\be
\label{gap}
-\langle\cA_0 f f\rangle \geq  a_0\langle f^2\rangle,\quad a_0>0,
\ee
for all functions $f,\langle f\rangle=0,$ in the domain of $\cA_0$.



The motion in this temporally stationary, Markovian flow  is 
also a temporally stationary, Markov process 
whose generator is
\be
\label{16'}
\cL=\cA+\widetilde{\vV}\cdot\nabla
\ee
when molecular diffusion is absent, and is
\be
\label{16''}
\cL=\cA+\kappa\Delta+\widetilde{\vV}\cdot\nabla
\ee
when molecular diffusion is present (\cite{markov}).

Now we make an observation which will be used later in
assessing the role of molecular diffusion.
The generator (\ref{16''}) in conjunction with (\ref{2.30}) and (\ref{gap})
suggests that
the presense of molecular
diffusion introduces a mechanism of generating
a Lagragian correlation in time comparable to
$\beta=1$ in the Eulerian correlation in time. 
For $\beta< 1$,
the generator $\cA$ dominates over $\kappa\Delta$
for low
wave numbers and, if a fixed ultraviolet cutoff is also present, are also
comparable to $\kappa\Delta$ for the other wave numbers.
Thus, the effect of molecular diffusion is negligible for $\beta\leq 1$
and $\alpha<1$ in the limit
of high Peclet number ($\kappa\to 0$).

In the sequel we shall use  the notation of 
the fractional gradiant of order $\beta$,
\[
\nabla^{\beta}:=(-\Delta)^{(\beta-1)/2}\nabla.
\]

\section{Transport properties of various wave numbers}
To study motion in a flow with a power-law energy spectrum
over a wide range of scales, it is convenient to 
decompose the energy spectrum into the 
the {\em sampling drift}  and 
the {\em fluctuating} velocity field,
and to consider
separately their distinctive transport properties. 
The relation between the sampling drift
and the fluctuating velocity field is
like that between a mean flow
and the fluctuation. 

\subsection{Sampling drift and critical wave numbers}
For each realization of random velocity field there is a nonzero
sampling drift due to random fluctuation,
depending on the scale of observation.

The volume-averaged flow on the observation scale
consists of 
spatially non-fluctuating wave numbers
on the observation scale, namely,
all
$|\bk|\sim \veps$. 
The the volume-averaged flow comprises three kinds of wave numbers:
{\em supercritical, critical} and {\em subcritical} wave numbers
depending
on their variations in time on the observation scale.
Critical and supercritical wave numbers compose the 
{\em sampling drift}.

The supercritical wave numbers are
effectively {\em
steady}
in the sense that
their correlation times are much larger than the time scale of observation,
$|\bk|^{-2\beta}\gg\ep^{-2q}$, thus, satisfy
\be
\label{16}
|\bk|\ll\min{\{\ep^{q/\beta},\ep\}}.
\ee
As these wave numbers are temporally as well as
spatially uniform, 
they behave like a constant drift on the observation scale and
transport particles {\em ballistically}. 
Among them,
we pay special attention to those wave numbers that,
on their own correlation time scales,
transport particles over a distance larger than
the observation scale
\beq
\label{super}
|\bk|^{1-\alpha}|\bk|^{-2\beta}
\gg 1/\ep,
\eeq
since
wave numbers of order $|\bk|$ have an amplitude of the order
\be
\label{am}
\left(\int_{c_1|\bk|\leq |\bk'|\leq c_2|\bk|} \cE(\bk')d|\bk'|\right)^{1/2}\sim
|\bk|^{1-\alpha},\quad |\bk|\ll 1.
\ee
Note that, for (\ref{super}) to define a non-empty set of low wave numbers,
we need
$\alpha+2\beta>1$.
For $\alpha+2\beta\leq 1$, the supercritical wave numbers
do not contribute to the transport on the observation scale
and are negligible asymptotically.
In any case, the transport effect of
the insignificant supercritical wave numbers are by nature negligible.

Since we do not know the scaling exponent $q$ ahead of time,
we define the {\em critical} wave numbers to be the
boundary of those {\em significant}
supercritical wave numbers. Thus, 
the critical wave numbers are of the order
$k_c=\ep^{\gamma_c}$, with
\be
\label{critical}
\gamma_c=\max{\{1, (\alpha+2\beta-1)^{-1}\}}=\left\{\begin{array}{ll}
(\alpha+2\beta-1)^{-1},&\hbox{for}\,\,1<\alpha+2\beta<2\\
1,&\hbox{for}\,\,\alpha+2\beta\geq 2.
\end{array}
\right.
\ee
By (\ref{critical}), for $\alpha+2\beta >2$, the
the 
sampling drift is identical to the volume-averaged flow.

Insignificant supercritical wave numbers occur when the following conditions
are satisfied: $\gamma_c>q/\beta>1$. 
This leads immediately to
$\alpha+2\beta<2$ and $\beta\leq 1$. The latter follows from 
$q\leq 1$ (see Section~4.1). As we will see later,
this can only happen, 
in part of Regime I (with $q=1$) defined by
$\alpha+\beta<1,\alpha+2\beta<2,\beta<1$.

The critical wave numbers have long-range correlation
in time {\em or} in space and dominate the transport in
the fractional-Brownian-motion regimes (Regimes II and III).
The {\em subcritical} wave numbers
are either temporally fluctuating,
$|\bk|\gg\ep^{q/\beta},$
{\em or} spatially fluctuating
$|\bk|\gg\ep.$
Effectively, the subcritical wave numbers can be defined
by $|\bk|\gg k_c$ and, by definition, include the insignificant supercritical
wave numbers.

Denote
by $\bmc_\ep$ the sampling drift
on scale $1/\ep$. It 
has an amplitude of the order
\be
\label{cs}
\left(\int_{\delta\leq |\bk|\leq k_c} \cE(\bk)d|\bk|\right)^{1/2}
\sim\left\{\begin{array}{ll}
\left|k_c^{1-\alpha}-\delta^{1-\alpha}\right|,&\hbox{for}\,\,\alpha\neq
1\\
\left|\log{k_c}-\log{\delta}\right|,&\hbox{for}\,\,\alpha=1.
\end{array}
\right.
\ee
Since the critical wave numbers dominate the sampling drift for
$\alpha<1$, $\bmc_\ep$ has a long-range correlation in space or time
on the observation scale, so 
its transport effect
is not ballistic. For $\alpha\geq 1$, $\bmc_\ep$ is effectively frozen
in time and its transport effect is ballistic.

Infrared cutoffs are classified accordingly: $\delta=\ep^\gamma$
is {\em critical} if $\gamma=\gamma_c$,
{\em supercritical} if $\gamma<\gamma_c$, and
{\em subcritical} if $\gamma>\gamma_c$.
We call $\gamma_c$ the {\em critical exponent}.

From (\ref{16}) and (\ref{super}), we have the simple inequality for
the scaling exponent
\be
\label{17}
q\leq \left\{\begin{array}{ll}
\beta/(\alpha+2\beta-1),& \hbox{for}\,\, 1<\alpha+2\beta < 2\\
\beta,& \hbox{for}\,\,\alpha+2\beta\geq 2
\end{array}
\right.
\ee
The equality in (\ref{17}) is admissible
because
asymptotics
is a continuum and can not be fully resolved by power-laws.
Here we restrict our attention to the power-law part of scaling behaviors.

For transport effect, besides the line $
\alpha+2\beta=2$, the line $\alpha=1$ is also important for
the following reasons. 
For $\alpha<1$,
the sampling drift is dominated by the critical wave numbers,
whereas,
for $\alpha\geq 1$, the sampling drift
is dominated by wave numbers nearby the infrared cutoff.
Moreover, in the case of $\alpha\geq 1$, the infrared cutoff
corresponds to the energy containing scale and, therefore, dominates the
the transport as well as the flow. As a result, scaling laws of
transport for
$\alpha\geq 1$ are in general (infrared) cutoff dependent.
The limit processes in the case of a {\em supercritical} cutoff, however,
are always regular motions ($H=1$, Regime IV)
as the effectively constant drift
dominates the transport.

Based on the supercritical wave numbers alone,
the exit time $\tau$ (out of a ball of radius $1/\epsilon$)
for $\alpha<1$
can be estimated by
\be
\label{exit}
\tau\ll k_c^{\alpha-1}/\ep=
\left\{\begin{array}{ll}
\ep^{-2\beta/(\alpha+2\beta-1)},&
\hbox{for}\,\, 1<\alpha+2\beta<2\\
\ep^{-2+\alpha},&\hbox{for}\,\, \alpha+2\beta\geq 2.
\end{array}
\right.
\ee
It is easy to see that, the (asymptotic)
equality in (\ref{exit}) is achieved when the combined effect of
the supercritical {\em and} the critical wave numbers is considered
since, for $\alpha<1$, the critical wave numbers are much stronger
than the supercritical wave numbers in magnitude. For $\alpha\geq 1$,
however, the transport is dominated by the wave numbers $|\bk|\sim\delta$.
So we have
\be
\label{exit2}
\tau\ll
\delta^{\alpha-1}/\ep=
\ep^{-1-\gamma+\alpha\gamma} \quad(\hbox{with $\delta=\ep^\gamma$})
\ee
in the case of $\alpha\geq 1$.

As we will show by the variational method in Section~3.3
that the critical wave numbers dominate
the transport in Regimes II and III.
In the case of $\alpha<1$, the exponent
for $\alpha+2\beta<2$ is less than
or equal to 2 (i.e.,
$2\beta/(\alpha+2\beta-1)\leq 2$) only if $\alpha+\beta\geq 1$;
for $\alpha+2\beta\geq 2$, the exponent is less than or equal to 2
(i.e., $2-\alpha<2$) only if $\alpha\geq 0$. The former defines Regime II;
the latter defines Regime III. In the case of
$\alpha\geq 1$, any nonnegative $\gamma$ leads to
$1+\gamma-\alpha\gamma\leq 2$ (the
scaling is superballistic for $\gamma>1$).
In the remaining region (Regime I: $\alpha+\beta<1$ or $\alpha<0$),
the supercritical wave numbers
are negligible
since the transport effect of the fluctuating wave numbers is at least
{\em diffusive} as we will see later.
Equating the exponent with $2q$,
we have, from (\ref{exit}), the scaling exponents for
Regimes II, III (see Section 4),
and, from eq. (\ref{exit2}), the scaling exponent
for Regime IV (see Section 5), both with {\em supercritical} cutoffs,
$\gamma>\gamma_c$.

In the regimes where the critical wave numbers
have a leading effect, the scaling limit is a fractional
Brownian motion (Regimes II and III).
Fractional Brownian motions arise as a result of long-range correlation
of the critical wave numbers.

\commentout{
Note that, to yield a superdiffusive
scaling, $\alpha+\beta$ must be larger than 1
in the first formula in (\ref{exit}), whereas 
$\alpha$ must be
larger than zero in the second
formula; the first condition defines the phase boundary between Regimes
II and I, whereas the second defines the phase boundary between
Regimes III and I. Thus, the critical sampling drift does not have a leading
order effect on transport in the diffusive Regime I (see
Section 4.1).
}

If the infrared cutoff is subcritical, i.e., $\delta\gg k_c$, wave numbers
of the spectrum are either temporally or spatially fluctuating.
Contrary to the fractional Brownian motion limit caused by the
critical sampling drift, the limit is always a Brownian motion.
But the scaling exponent may be superdiffusive due to low wave numbers
in the vicinity of the cutoff.

\subsection{Subcritical wave numbers:
eddy diffusivity}
To study the effect of subcritical, or fluctuating, wave numbers on transport,
we think of turbulent motion 
as a superposition of a mean flow (i.e. $\bmc_\ep$),
and the fluctuating flow,
following a spectral discretization.

We propose that 
the fluctuating wave numbers give rise to a fluctuating motion, 
on top of the mean flow,
on the observation scale, and this fluctuating motion can be characterized
by a 
notion
of scale dependent eddy diffusivity introduced below.
We then formulate
two variational principles and use them to obtain  general
upper bounds for the (scale-dependent) eddy diffusivity. 

Spectral discretization is motivated
by a standard result of the ergodic theory for stationary processes that
a stationary process is a limit
of periodic processes (see \cite{D}).
We will use the periodic approximation
in two different ways: In the first, we consider the periodic approximation
in the {\em space} variables only and work with
a subspace of $(\Om,P)$,
the space $(\Om^{(n)},P^{(n)})$ of
time-independent, space-periodic velocity fields with period cell $[0,n]^d$
(see discussion below).
In this approach, time randomness in the velocity field
is represented as a Markov process on $(\Om^{(n)},P^{(n)})$.
In the second approach,
we work with a sequence of {\em space-time} periodic fields
with the (normalized) Lebesgue measure as the probability distribution
on the space-time period cells as stated in the following lemma.
\begin{lemma}
\label{lemma1}
Let $\om$ be a stationary process. Then there exists
a sequence of periodic processes $\om_n$ of period $\ell_n\to \infty$ in
each variable, such that, the probability measure $P_n$ obtained
as the distribution of $\tau_{\bx}\om_n$ where $\bx$ is random and distributed
uniformly on the period cell $[0,\ell_n]^d$ converges weakly to the distribution of $\om$ as
$n\to\infty$.
\end{lemma}
(See, for instance, \cite{Pa} for a proof).
We emphasize that spectral discretization is only a convenience 
for the formulation of the variational principles; it is neither essential 
nor necessary. 

We now formulate the first approach more specifically.
A spectrally discretized flow can be written as a sum 
$\bmc_\ep +\vV^{(\ep,n)}
$
where $\bmc_\ep$ is the sampling drift
(see Section 3.1 and eq. (\ref{cs})), and 
$\vV^{(\ep,n)}$ is the {\em spatially periodic}
version of the fluctuating velocity field
with a discrete spectrum $\bk\in Z^d/n, \max{\{k_c,\delta\}}\ll |\bk|\leq K$
and the amplitude
$(\bI-\bk\otimes\bk/|\bk|^2)
\sqrt{\int_{|\bk|\leq |\bk'|\leq |\bk|+1/n} \cE(\bk')d|\bk'|}
$. 
The mesh size $1/n$ should tend to zero sufficiently fast, as $\ep\to 0$,
to approximate the transport effect of the original fluctuating
flow in view of the above lemma.

Equivalently,
we replace the spectral measure $\widehat{\vV}(d\bp,t)$
by the discrete measure $\widehat{\vV}^{(\ep,n)}(\bk,t)\delta_{\bk,\bp},$
$\forall
\bk\in Z^d/n, \max{\{k_c,\delta\}}\ll |\bk|\leq K$ with 
$\widehat{\vV}^{(\ep,n)}(\bk,t)$ 
satisfying
\be
\label{12.30'}
d_t\widehat{\vV}^{(\ep,n)}(\bk,t)=-a_0|\bk|^{2\beta}
\widehat{\vV}^{(\ep,n)}(\bk,t)dt+|\bk|^\beta
\sqrt{\int_{|\bk|\leq |\bk'|\leq |\bk|+1/n} \cE(\bk')d|\bk'|}
\left(\bI-\bk\otimes\bk|\bk|^{-2}\right)^{1/2}d_t\bW(
\bk,t)
\ee
where $\bW(\bk,t),\forall\bk\in  Z^d/n, \max{\{k_c,\delta\}}\ll
|\bk|\leq K$ are independent standard Brownian motions.
As discussed in Section 3.1, the sampling drift $\bmc_\ep$
is steady for $\alpha+2\beta>2$ or $\alpha\geq 1$; it is unsteady
for $\alpha+2\beta\leq 2, \alpha<1$.

The time-stationary, space-periodic field
$\vV^{(\ep,n)}(\bx,t,\om_n),\om\in\Om^{(n)}$
is a Markovian flow  and can be
represented as a translate,
$\vV^{(\ep,n)}(\bx,t,\om_n)=\widetilde{\vV}^{(\ep,n)}(\bx,\om_n(t))$,
of steady, space-periodic field
$\widetilde{\vV}^{(\ep,n)}(\bx,\om_n)$ where
$\om_n(t), w_n(0)=0$ is a Markov process  on 
$\Om^{(n)}$.
As usual, we write $\om_n$ explicitly only to emphasize its role.

\commentout{
The generator of $\vV^{(\ep,n)}(\bx,t)$
is $\cA^{(\ep,n)}=(-(4\pi^2)^{-1}\Delta)^\beta\cA_0^{(\ep,n)}$
where $\cA_0^{(\ep,n)}$ is the generator of the process
$\vV^{(\ep,n)}_0$ satisfying
\[
d_t\widehat{\vV}_0^{(\ep,n)}(\bk,t)=-a_0
\widehat{\vV}_0^{(\ep,n)}(\bk,t)dt+
\sqrt{\int_{|\bk|\leq |\bk'|\leq |\bk|+1/n} \cE(\bk')d|\bk'|}
\left(\bI-\bk\otimes\bk|\bk|^{-2}\right)^{1/2}d_t\bW(
\bk,t),
\]
$\forall \bk\in  Z^d/n, \max{\{k_c,\delta\}}\ll |\bk|\leq K$.
The generator for $\bmc^\ep (t)+\vV^{(\ep,n)}(\bx,t)$
is $\cA^{(\ep,n)}=(-(4\pi^2)^{-1}\Delta)^\beta\cA_0^{(\ep,n)}$
where $\cA_0^{(\ep,n)}$ is the generator of the process
$\vV^{(\ep,n)}_0$ satisfying
\[
d_t\widehat{\vV}_0^{(\ep,n)}(\bk,t)=-a_0
\widehat{\vV}_0^{(\ep,n)}(\bk,t)dt+
\sqrt{\int_{|\bk|\leq |\bk'|\leq |\bk|+1/n} \cE(\bk')d|\bk'|}
\left(\bI-\bk\otimes\bk|\bk|^{-2}\right)^{1/2}d_t\bW(
\bk,t),
\]
$\forall \bk\in [\delta,K]\cap Z^d/n, |\bk|\gg \max{\{k_c,\delta\}}$,
and $\cC_0^{(n)}$ the generator of
\[
d_t\bmc_0^\ep(t)=\left\{\begin{array}{ll}
-a_0\bmc_0^\ep(t) dt+
\sqrt{\int_{\delta\leq |\bk'|\leq 1/n} \cE(\bk')d|\bk'|}
d\bW(0,t),&
\hbox{if $\alpha+2\beta\leq 2, \alpha <1$}\\
0,&\hbox{if else}
\end{array}\right.
\]
}

For fixed $\ep, n$, the displacement, $\bmx(t)$, in the periodic flow,
$\bmc_\ep +\vV^{(\ep,n)}$, is the sum of a mean motion,
$\int^t_0\bmc_\ep(s) ds$,
and the fluctuation, $\bmx(t)-\int^t_0\bmc_\ep (s) ds$.
After a proper rescaling $t\to \lambda^2 t, \bx\to\lambda\bx,\lambda\to \infty$,
the fluctuation converges to a Brownian motion 
by a turbulent diffusion theorem for mixing flows (\cite{markov}).
Let $\cA^{(\ep,n)}$ be 
the generator for $\bmc^\ep (t)+\vV^{(\ep,n)}(\bx,t)$.
The diffusion coefficients, $D_{ij}^{(\ep, n)}$,
of the limiting Brownian motion
are determined from the random, {\em space-periodic} solution
$\chi_j^{(\ep, n)}$ (i.e. $\chi_j^{(\ep,n)}$ can be viewed as a function
defined on $\Om^{(n)}$) of the abstract cell problem (cf. (\ref{16'}),
see also \cite{markov})
\beq
\label{2.13}
&\cL^{(\ep,n)}\chi^{(\ep,n)}_j&:=\cA^{(\ep,n)}\chi^{(\ep,n)}_j +\left(\bmc_\ep +\widetilde{\vV}^{(\ep,n)}\right)
\cdot\nabla\chi^{(\ep,n)}_j=-\widetilde{V}^{(\ep,n)}_j
,\quad \hbox{in}\quad\Om^{(n)}, \quad\forall i,j\\
\nonumber 
&D^{(\ep,n)}_{ij}&:={1\over 2}\left(\langle\widetilde{V}^{(\ep,n)}_j\chi^{(\ep,n)}_i\rangle_n
+\langle\widetilde{V}^{(\ep,n)}_i\chi^{(\ep,n)}_j\rangle_n\right)\\
&&=-{1\over 2}\left(\langle \cL^{(\ep,n)}
\chi^{(\ep,n)}_i \chi^{(\ep,n)}_j\rangle_n+
\langle \cL^{(\ep,n)}\chi^{(\ep,n)}_j \chi^{(\ep,n)}_i\rangle_n\right)\nonumber\\
&&=-{1\over 2}\left(\langle \cA^{(\ep,n)}
\chi^{(\ep,n)}_i \chi^{(\ep,n)}_j\rangle_n+\langle \cA^{(\ep,n)}
\chi^{(\ep,n)}_j \chi^{(\ep,n)}_i\rangle_n\right)\nonumber\\
&&=
\langle \nabla^{\beta}\chi_i^{(\ep,n)}
\cdot\cA^{(\ep,n)}_0\nabla^\beta\chi^{(\ep,n)}_j\rangle_n,
\quad\forall i,j. \label{2.14}
\eeq
with the periodic boundary condition,
where $\langle\cdot\rangle_n$ is the average with respect to $P^{(n)}$.
Here we have used the following identity 
\beq
&&\langle\left[(\bmc_\ep +\widetilde{\vV}^{(\ep,n)})\cdot\nabla\chi^{(\ep,n)}_i\right]
\chi^{(\ep,n)}_j\rangle_n+
\langle\left[(\bmc_\ep +\widetilde{\vV}^{(\ep,n)})\cdot\nabla\chi^{(\ep,n)}_j\right]
\chi^{(\ep,n)}_i\rangle_n\nonumber\\
&&=
\nabla\cdot\langle(\bmc_\ep +\widetilde{\vV}^{(\ep,n)})
\chi^{(\ep,n)}_i\chi^{(\ep,n)}_j\rangle_n =0
\eeq
which follows from the incompressibility of $\bmc_\ep +\widetilde{\vV}^{(\ep,n)}$
and the space-homogeneity of  $
\langle(\bmc_\ep +\widetilde{\vV}^{(\ep,n)})
\chi^{(\ep,n)}_i\chi^{(\ep,n)}_j\rangle_n$.

The problem (\ref{2.13}) is well-posed and has
a unique solution, up to a constant (which does not affect
(\ref{2.14})).
In Appendix, we derive the minimum principle
\be
\label{2.22}
D^{(\ep,n)}(\bfe):= \bD^{(\ep,n)}\bfe\cdot\bfe
=\inf_{f}
\left\{-\langle\cA^{(\ep,n)} f f\rangle_n-\langle \cA^{(\ep,n)}
f'f'\rangle_n\right\}
\ee
with the space-periodic functions $f',f$ related by
\be
\label{2.25}
\cA^{(\ep,n)} f'+(\bmc_\ep +\widetilde{\vV}^{(\ep,n)})
\cdot\nabla f+\widetilde{\vV}^{(\ep,n)}\cdot\bfe=0,
\quad \hbox{in}\quad\Om^{(n)}.
\ee
It should be noted that the explicit form of the generator
is not used for the variational formulation.

In the limit $n\to\infty$, the abstract cell problem
(\ref{2.13})-(\ref{2.14}) becomes
\be
\label{2.8'}
\cL\chi^{(\ep)}_j+\widetilde{V_j}^{(\ep)}=0,\quad\hbox{in}\,\,\Om.
\ee
We also have
\beq
D^{(\ep)}_{ij}&=&\lim_{n\to\infty}D^{(\ep,n)}_{ij}\nonumber\\
&=&
{1\over 2}\left(\langle\widetilde{V}^{(\ep)}_j\chi^{(\ep
)}_i\rangle
+\langle\widetilde{V}^{(\ep)}_i\chi^{(\ep)}_j\rangle\right)\nonumber\\
&=&
\langle \nabla^{\beta}\chi_i^{(\ep)}
\cdot\cA_0\nabla^\beta\chi^{(\ep)}_j\rangle,
\quad\forall i,j,
\label{2.14'}
\eeq
following from  (\ref{2.14}).
It is clear from (\ref{2.14'}) 
that the matrix $\bD^{(\ep)}=[D^{(\ep)}_{ij}]$ is symmetric and
positive-definite.
We think of $\bD^{(\ep)}$ as a measure of
turbulent dispersion caused by eddies composed of
subcritical wave numbers in interaction with the sampling drift.
We call it
the {\em
eddy diffusivity}.
If
the increments of
the fluctuation of particle motion have
divergent step sizes as $\ep\to 0$, 
then
the eddy diffusivity is cutoff-dependent. 
Eq. (\ref{2.14'}) 
indicates the right solution
space for (\ref{2.8'}): $L^2_\beta(\Om)$,
the space
of functions with homogeneous, square integrable
fractional gradient of order $\beta$.

\commentout{
{\em If} the abstract cell problem (\ref{2.13})-(\ref{2.14}) converges
as $n$ tends to infinity, then the limit problem should be 
\be
\label{2.8}
\cL\chi_j+\widetilde{V_j}=0,\quad\hbox{in}\,\,\Om
\ee
and the diffusion coefficients should converge to
$\langle \nabla^{\beta}\chi_i
\cdot\cA_0\nabla^\beta\chi_j\rangle.$
This expression
is nonnegative definite and indicates the right solution
space for (\ref{2.8}): $L^2_\beta(\Om)$,
the space
of functions with homogeneous, square integrable
fractional gradient of order $\beta$.
}

From another perspective, the variance of the fluctuation
$\bmx(t)-\int^t_0\bmc_\ep (s) ds$ 
after the rescaling $t\to\lambda^2 t,\bx\to\lambda\bx, \lambda\to\infty$
can be expressed as the time integral
\[
2\int^t_0 D^{\ep,\lambda}_{ij}(s) ds
\]
of the Lagrangian velocity autocorrelation 
\be
D_{ij}^{\ep,\lambda}(s)={1\over 2}\int^{\lambda^2s}_0\left(
\langle V^{(\ep)}_i (\bmx(s),
s) V^{(\ep)}_j (\bmx(s'),s')+
\langle V^{(\ep)}_j (\bmx(s),
s) V^{(\ep)}_i (\bmx(s'),s')\rangle\right)
ds'.
\label{2.10}
\ee
Because the Lagrangian velocity
$\vV^{(\ep)}(\bmx(t),t)
$
is a stationary
Markov process
(\cite{PS},\cite{markov}),
(\ref{2.10})
can be rewritten as
\beq
\nonumber
D_{ij}^{\ep,\lambda}(s)
&=&{1\over 2}
\int^{\lambda^2s}_0\left\{\langle V^{(\ep)}_i(0,0) V^{(\ep)}_j 
(\bmx(s'),s')\rangle
+\langle V^{(\ep)}_j(0,
0) V^{(\ep)}_i (\bmx(s'),s')\rangle\right\}\,\,ds'\\
&=&{1\over 2}
\int^{\lamb^2s}_0 \left\{\langle \widetilde V^{(\ep)}_i \exp{(\cL
s')}\widetilde V^{(\ep)}_j\rangle
+\langle \widetilde V^{(\ep)}_j \exp{(\cL s')}\widetilde V^{(\ep)}_i\rangle\right\}\,\,ds'.
\nonumber
\eeq
In the limit $\lambda\to \infty$,
$D_{ij}^{\ep,\lambda}(s)$ tends to
the following expressions
\be
\label{2.7}
-{1\over 2}\left(\langle \widetilde{V_i}^{(\ep)}\cL^{-1}
\widetilde{V_j}^{(\ep)}\rangle+
\langle \widetilde{V_j}^{(\ep)}\cL^{-1}\widetilde{V_i}^{(\ep)}\rangle\right)
={1\over 2}\left(
\langle \widetilde{V_i}^{(\ep)}\chi^{(\ep)}_j\rangle+\langle \widetilde{V_j}^{(\ep)}\chi^{(\ep)}_i\rangle
\right)
=D^{(\ep)}_{ij}
\ee
where $\chi^{(\ep)}_j$ is the solution of (\ref{2.8'}).

When molecular diffusion is present, 
we denote
the eddy diffusivity 
by $\bD^{(\ep,n)}_\kappa$. As before, $\bD^{(\ep,n)}_\kappa$
can be
characterized
variationally by adding the terms, $\kappa, \kappa
\langle \nabla f\cdot\nabla f\rangle_n,
\kappa\langle \nabla f'\cdot\nabla f'\rangle_n$ to
(\ref{2.22}) and a Laplacian term, $\kappa\Delta f'$, to
(\ref{2.25}). 

We turn to the second approach of space-time periodic approximation.
Let
$\vV^{(\ep,n,\lambda)}(\bmx,t)$ be
the approximating
sequence of space-time periodic fields, as stated in Lemma~\ref{lemma1},
with increasing space period $n$ and time period $\lambda$, for
the velocity field $\vV^{(\ep)}$ with a subcritical cutoff.
We work with the space-time period cell problem in which
time randomness in the Lagrangian dynamics is absent. To formulate
a variational principle in this case, we need to reinstate the
molecular diffusion here.

For fixed $\lambda,n$, the effective diffusivity, $\bD^{(\ep,n,\lambda)}_\kappa$,
in the flow
$\vV^{(\ep,n,\lambda)}(\bmx,t)$ exists and can be given as
\be
\label{eff}
D^{(\ep,n,\lambda)}_\kappa(\bfe)=\inf_{f} {1\over \lambda}\int^\lambda_{0}
{1\over n^d}\int_{[0,n]^d}
\kappa(1+\nabla f\cdot\nabla f+\nabla f'\cdot
\nabla f')\,\,d\bx\,\,dt
\ee
where $f,f'$ are both temporally and
spatially periodic with the period $\lambda,n$, respectively,
and are related to each other by the
following equation
\[
\kappa\Delta f'=-{\partial f\over \partial t}-(\bmc_\ep +\vV^{(\ep,n,\lambda)})
\cdot\nabla f-
\vV^{(\ep,n,\lambda)}\cdot\bfe
\]
(see \cite{ced} and \cite{dt}).
The eddy diffusivity $\bD^{(\ep)}_\kappa$ 
in the presence of molecular diffusion
is the large scale limit of $D^{(\ep,n,\lambda)}_\kappa$, i.e.,
\be
\label{eff'}
\bD^{(\ep)}_\kappa=\lim_{n,\lambda\to \infty}
\bD^{(\ep,n,\lambda)}_\kappa.
\ee

The variational principle (\ref{eff})-(\ref{eff'})
is more useful than (\ref{2.22})-(\ref{2.25}) when
the Laplacian in the generator (\ref{16''})
dominates $\cA$ for {\em low} wave numbers as in the case of 
$\beta>1$. 
Another advantage for working with the space-time periodic setting
is that 
a dual variational principle can be formulated  for the inverse of
$D^{(\ep,n,\lambda)}_\kappa$ 
and can be used to obtain the lower bound for
$D^{(\ep,n,\lambda)}_\kappa$ (see \cite{ced},\cite{rced},\cite{an}).


\subsection{Variational bounds: fractional vector potential}
\subsubsection{Case 1: Supercritical cutoff}
When the sampling drift is present, i.e., $\gamma>
\gamma_c$,  we show
by the variational principles the following
upper bounds on the growth rate of the eddy diffusivity
\beq
D_{ij}^{(\ep)}&\leq& C,\quad\hbox{for $\alpha+\beta< 1$ {\em or} $\alpha<0$}
\nonumber\\
\label{upper}
D_{ij}^{(\ep)}&\ll&
\left\{
\begin{array}{ll}
\ep^{2\gamma_c(1-\alpha-\beta)},&\hbox{for $\alpha+\beta>1$}\\
\log{(1/\ep^{\gamma_c})},& \hbox{for $\alpha+\beta= 1$}\\
\ep^{-\alpha},& \hbox{for $\alpha>0$}\\
\sqrt{\log{(1/\ep^{\gamma_c})}},& \hbox{for $\alpha=0$}\\
\end{array}
\right.,
\eeq
$\quad\forall i,j$, for some constant $C>0$.
Note that $\ep^{-\alpha}$ is a better bound than
$\ep^{2\gamma_c(1-\alpha-\beta)}$ for $\alpha+2\beta\geq 2$.

Take the trivial trial function $f=0$ in (\ref{2.22}) and eliminate
the first term in the functional. We calculate the second term in (\ref{2.22})
by studying the equation
\be
\label{2.31}
\cA^{(\ep,n)} f'+\vV^{(\ep,n)}\cdot\bfe=0
\ee
(cf. (\ref{2.25})).
Consider the {\em fractional} vector potential (the fractional
stream function) in three dimensions (in two dimensions)
$\widetilde{\bH}^{(\ep,n)}_{\beta}$ of order $\beta$ defined by 
\be
\label{pot}
\widetilde{\bH}^{(\ep,n)}_\beta=(-\Delta)^{-\beta/2}\widetilde{\vV}^{(\ep,n)}
\ee
via the Fourier transform.
This means that $
\widetilde{\bH}^{(\ep)}_\beta=\lim_{n\to \infty}
\widetilde{\bH}^{(\ep,n)}_\beta$
has the energy spectrum
\be
\label{2.33}
E_0|\bk|^{1-2(\alpha+\beta)}
\ee
with a subcritical infrared cutoff.
The usual vector potential and the stream function correspond to
$\beta=1$.  
What is significant is that, for $\alpha+\beta<1$,
(\ref{2.33}) is integrable near
$\bk=0$ uniformly as the the infrared cutoff is removed,
and thus defines a homogeneous, square integrable $\bH_\beta$
that is cutoff-independent.
For $\alpha+\beta\geq 1$, (\ref{2.33}) is not square integrable
uniformly as the the infrared cutoff is removed, and  
the second moment of $\bH^{(\ep)}_{\beta}$
grows like
\be
\label{26'}
\langle|\bH^{(\ep)}_{\beta}|^2\rangle
\ll\ep^{2\gamma_c(1-\alpha-\beta)},\quad \hbox{for}\,\, \alpha+\beta>1
\ee
and
$\langle|\bH^{(\ep)}_{\beta}|^2\rangle \ll\log{(1/\ep^{\gamma_c})}$,
for $\alpha+\beta=1$
as $\ep\to 0$.

In terms of $\bH^{(\ep,n)}_{\beta}$,  (\ref{2.31}) becomes
\be
\label{2.31'}
\cA^{(\ep,n)} f'+(-\Delta)^{\beta/2}\bH^{(\ep,n)}_{\beta}\cdot\bfe=0.
\ee
A straightforward energy estimate for (\ref{2.31'}) gives
\beq
-\langle\cA^{(\ep,n)} f'f'\rangle_n&=&\langle(-\Delta)^{\beta/2}\bH^{(\ep,n)}_{\beta}\cdot\bfe f'\rangle_n\\
&\leq& \sqrt{\langle|\bH^{(\ep,n)}_{\beta}\cdot\bfe|^2\rangle_n}\sqrt{\langle|(-\Delta)^{\beta/2}f'|^2\rangle_n}\\
&\leq&\sqrt{\langle|\bH^{(\ep,n)}_{\beta}\cdot\bfe|^2\rangle_n}\sqrt{
{1\over a_0}\langle-\cA^{(\ep,n)} f'f'\rangle_n}.
\eeq
Therefore
\be
\label{35}
-\langle\cA^{(\ep,n)} f'f'\rangle_n\leq
{1\over a_0}
\langle|\bH^{(\ep,n)}_{\beta}\cdot\bfe|^2\rangle_n
\ee
which, in the limit $n\to\infty$, is much less than $ \ep^{2\gamma_c
(1-\alpha-\beta)}$ for small $\ep$.

It is worth noting
that the right side of (\ref{35}) is, up to a factor independent
of $\ep$, what one gets
in replacing the Lagrangian autocorrelation in 
(\ref{2.10}) by the Eulerian autocorrelation $R^{(\ep)}_{ij}(0,s-s'):=
\langle V^{(\ep)}_i(0,s)V^{(\ep)}_j(0,s')\rangle$, i.e.
\be
\label{TK'}
\int^{\infty}_0 R^{(\ep)}_{ij}(0,s)
ds
\ee
which is called the (Eulerian) Taylor-Kubo formula,
used extensively in the literature to approximate the 
eddy diffusivity since Taylor's work (\cite{T}).
The physical significance of the bound (\ref{35}) is that
{\em 
the eddy diffusivity $D^{(\ep)}_{ij}$ of the fluctuation
is bounded, as the infrared cutoff is removed, 
by 
a constant times
the Eulerian Taylor-Kubo formula (\ref{TK'})};
the eddy diffusivity may be much smaller than 
(\ref{TK'}) due to
the spatial decorrelation of velocity.


A different upper bound for the eddy diffusivity
the can be obtained by using the second
variational principle  (\ref{eff}).
First we note that
the eddy diffusivity in the presence of
molecular diffusion would be enhanced if
we freeze the time variable
of the velocity field $\vV^{(\ep)}$.
This can be easily seen as follows.
Let $\bar{\bD}^{(\ep,n)}_\kappa$
be the eddy diffusivity 
for the frozen velocity field $\vV^{(\ep,n)}(\bx,0)$
\be
\label{frozen}
\bar{D}^{(\ep,n)}_\kappa(\bfe):=
\bar{\bD}^{(\ep,n)}_\kappa\bfe\cdot\bfe=\inf_{f}
{1\over n^d}\int_{[0,n]^d}\kappa(1+\nabla f\cdot\nabla f+\nabla f'\cdot
\nabla f')\,\,d{\bf x}
\ee
where $f, f'$ are spatially periodic with period cell $[0,n]^d$
and are related by
\[
\kappa\Delta f'=-(\bmc_\ep +\vV^{(\ep,n)}(\bmx,0))\cdot\nabla f-
\vV^{(\ep,n)}(\bmx,0)\cdot\bfe.
\]
Since time independent trial functions $f$ are admissible 
in (\ref{eff}), 
(\ref{frozen}) is larger than $D^{(\ep,n,\lambda)}_\kappa, \forall \lambda>0,$
given by (\ref{eff}).

Using the trivial trial function $f=0$ in 
(\ref{frozen}) and do the same energy estimate as above
we have the upper bound
\[
\bar{\bD}^{(\ep)}_\kappa:=
\lim_{n\to \infty}\bar{\bD}^{(\ep,n)}_\kappa 
\leq\langle|\bH^{(\ep)}_{1}|^2\rangle.
\]
A better bound, however, can be obtained for
steady, isotropic flows by a duality argument
in conjunction with the variational method (see (\cite{an}):
\be
\label{up}
\bar{\bD}^{(\ep)}_\kappa\leq
\sqrt{\langle|\bH^{(\ep)}_{1}|^2\rangle}
\left\{
\begin{array}{ll}
\leq C,&\hbox{for $\alpha<0$}\\
\ll \ep^{-\alpha},&\hbox{for $\alpha>0$}\\
\ll\sqrt{\log{(1/\ep^{\gamma_c})}},&\hbox{for $\alpha=0$}
\end{array}
\right.
\ee
which 
agrees with results by other
approaches (such as RNG calculation of \cite{BG} and Green's function
method of \cite{KB2}).

When the sampling drift is present, the estimates
(\ref{upper}) can be used to compare the
transport effects of the sampling drift and the subcritical
wave numbers. For $\alpha+\beta>1$,
we have from (\ref{upper}) the bound for
the time scale of the fluctuation of particle motion
\[
\ep^{-2}/D^{(\ep)}(\bfe)\gg\ep^{-2}\ep^{2\gamma_c(\alpha+\beta-1)}
=\ep^{-2\beta/(\alpha+2\beta-1)},
\]
which, in Regime II,  is
the time scale of observation
as determined from the
sampling drift alone (cf. (\ref{exit})).
Therefore, the transport
in Regime II is dominated by the sampling drift.

For $\alpha+2\beta\geq 2$, 
(\ref{up}) implies the bound for
the time scale  associated with
the fluctuation of particle motion 
\[
\ep^{-2}/D^{(\ep)}(\bfe)\gg
\ep^{-2}
\ep^{\alpha}=\ep^{-2(1-\alpha/2)},
\]
which, in Regime III, is the time scale of observation as determined
from the sampling drift alone (cf. (\ref{exit})).
Thus, again, the transport
in Regime III is dominated by the sampling drift.

\subsubsection{Case 2: Subcritical cutoff}
When the sampling drift is absent, i.e.,
$\gamma<\gamma_c$, or negligible, instead of (\ref{upper}), we have
\beq
D_{ij}^{(\ep)}&\leq& C,\quad\hbox{for $\alpha+\beta< 1$ {\em or} $\alpha<0$}
\nonumber\\
\label{upper2}
D_{ij}^{(\ep)}&\leq&
\left\{
\begin{array}{ll}
C\ep^{2\gamma(1-\alpha-\beta)},&\hbox{for $\alpha+\beta>1$}\\
C\log{(1/\ep^{\gamma})},& \hbox{for $\alpha+\beta= 1$}\\
C\ep^{-\gamma\alpha},& \hbox{for $\alpha>0$}\\
C\sqrt{\log{(1/\ep^{\gamma})}},& \hbox{for $\alpha=0$}\\
\end{array}
\right.,
\eeq
$\quad\forall i,j$,
for some constant $C>0$.
Note again that $\ep^{-\gamma\alpha}$ is a better bound than
$\ep^{2\gamma(1-\alpha-\beta)}$ for $\alpha+2\beta>2$.

The estimates (\ref{upper2}) are derived by the same energy estimate
as before.
In this case, $\bmc_\ep=0$ and the velocity field consists entirely of
the subcritical wave numbers. As a result, $\ll$ in (\ref{upper})
becomes $\leq$ in (\ref{upper2}).

When the sampling drift is absent, the estimates (\ref{upper2})
yield a lower bound for the scaling exponent.
For $\alpha+\beta>1, \gamma<\gamma_c$
the fluctuation of particle motion is $1/\ep$ and is much less than
\[
\sqrt{\ep^{2\gamma_c(1-\alpha-\beta)}\ep^{-2q}}=
\ep^{\gamma_c(1-\alpha-\beta)-q}.
\]
Thus, we have
\be
\label{j21.1}
q\geq 1+\gamma-\gamma(\alpha+\beta),\quad\hbox{for}\,\,\alpha+\beta>1.
\ee
The bound (\ref{j21.1}) is sharp when temporal
fluctuations of the velocity fields are the dominant mechanism for transport
as in Regime V.

For $\alpha+2\beta\geq 2, \gamma<\gamma_c$,
(\ref{upper2}) implies
\[
\ep^{-1}\leq\sqrt{\ep^{-\gamma\alpha}\ep^{-2q}}=\ep^{-q-\gamma\alpha/2}.
\]
Thus, we have
\be
\label{VI}
q\geq 1-\gamma\alpha/2, \quad\hbox{for}\,\,\alpha+2\beta\geq 2,
\ee
which turns out to be sharp in Regime VI.

In the case of subcritical infrared cutoffs, it is often
useful to know if the wave numbers $|\bk|\sim\delta$ dominate
the transport or not. 
For this purpose, we modify the previous variational method 
to estimate the transport
effect of the wave numbers much larger than the subcritical infrared
cutoff. We replace $\bmc_\ep$ by the velocity field $\vU_\ep$ consisting
entirely of wave numbers $|\bk|\sim
\delta=\ep^\gamma$ and $\vV^{(\ep)}$
by the velocity field consisting of the
wave numbers $|\bk|\gg \ep^\gamma$. After
other corresponding modifications are made,
the variational method and the subsequent
energy estimate after substitution of the trivial trial function work
the same way.
We get the upper bound for the contribution, denoted by $\dot{\bD}^{(\ep)}=
[\dot{D}_{ij}^{(\ep)}]$,
of wave numbers
$|\bk|\gg \ep^\gamma$ to the total eddy diffusivity $D_{ij}^{(\ep)}$: 
\beq
\dot{D}_{ij}^{(\ep)}&\leq& C,\quad\hbox{for $\alpha+\beta< 1$ {\em or} $\alpha<0$}
\nonumber\\
\label{upper2'}
\dot{D}_{ij}^{(\ep)}&\ll&
\left\{
\begin{array}{ll}
\ep^{2\gamma(1-\alpha-\beta)},&\hbox{for $\alpha+\beta>1$}\\
\log{(1/\ep^{\gamma})},& \hbox{for $\alpha+\beta= 1$}\\
\ep^{-\gamma\alpha},& \hbox{for $\alpha>0$}\\
\sqrt{\log{(1/\ep^{\gamma})}},& \hbox{for $\alpha=0$}\\
\end{array}
\right.,
\eeq
$\quad\forall i,j$,
for some constant $C>0$.
For specific application of these bounds, see discussions for Regimes V and
VI.

\section{Phase diagram for $\alpha<1$}

\subsection{Regime I: diffusive limit}

First of all, as discussed in Section 3.1,
the sampling drift is negligible in this regime: In (\ref{exit}),
if $\alpha+\beta<1$ and $\alpha+2\beta <2$, then $2\beta/(\alpha+2\beta-2)>2$;
if $\alpha<0$ and $\alpha+2\beta \geq 2$, then $2-\alpha >2$. In either case,
the transport would be dominated by the fluctuating wave numbers.

When $\alpha+\beta<1$, (\ref{upper}) implies
\be
\label{33}
0\leq \liminf_{\ep\to 0} D^{(\ep)}(\bfe) \leq \limsup_{\ep\to 0}
 D^{(\ep)}(\bfe)<\infty.
\ee
As we will see below that $D^{(\ep)}(\bfe)$ should not vanish
in the limit,
so the scaling is diffusive $q=1$ and the scaling limit
should be a Brownian motion ($H=1/2$).
The limit $D^*(\bfe)=\lim_{\ep\to 0}D^{(\ep)}(\bfe)$, if exists,
is the (scale independent) eddy diffusivity.
Similarly, for $\alpha<0$, the variational bound (\ref{up})
implies the diffusive scaling limit.

The eddy diffusivity probably does not vanish in the limit  for the
following reason.
From the turbulent diffusion theorem for {\em mixing} flows, proved
in \cite{markov},
we know that, for $\beta=0, \alpha <1$, 
the scaling is diffusive ($q=1$)  and  the limit is a Brownian motion.
As $\beta$ increases, the velocity correlation
in time increases and so should the rate of transport. But the upper
bound (\ref{33}) for the eddy diffusivity tells us that it can not
enhance transport to the extent of changing the scaling limit
as long as $\alpha +\beta < 1$
(This scenerio has been rigorously justified 
in the region $\alpha<0,\beta\leq 1$ in a 
different turbulent diffusion theorem for {\em non-mixing} flows, proved
in \cite{markov}.).

For $\alpha<0$, the (ordinary) vector
potentials for the flows are 
time-stationary,
space-homogeneous and have finite moments.
Then the diffusion limit theorem of \cite{moser2} holds for 
such flows
if molecular diffusion is present (i.e. $q=1, H=1/2$ if $\kappa>0$). 
And the effective diffusivity  can be determined  from
a pair of variational principles (\cite{dt}), one of which is
(\ref{eff}).
This is manifest in the
existence of (ordinary) vector potentials when $\alpha<0$.
As shown in \cite{dt}, for {\em steady flows}, the existence of
space-homogeneous (ordinary) vector potentials is the  {\em sharp}
condition for a diffusive  scaling limit with molecular
diffusion.
As time dependence of velocity becomes important with $\beta\leq 1$, 
the phase boundary defined by  $\alpha+\beta=1,
\beta\leq 1$ points to the fact that 
the existence of  space-homogeneous, {\em fractional}
vector potentials becomes
the criterion for the diffusive scaling limit.



For $\alpha+\beta>1,\alpha\leq 0$
(thus, $\beta>1$ and $\alpha+2\beta>2$), both the 
sampling drift and high wave numbers are negligible.
The effect of molecular diffusion
may not be negligible since the Laplacian in
the Lagrangian generator (\ref{16''}) 
dominates over $\cA$ in {\em low} wave numbers.


\subsection{Regime II: space-freezing property}
As we have seen from the analysis of the sampling
drift and the applications of variational bounds,
for 
\[
\gamma>\gamma_c,\quad \alpha+\beta>1,\quad \alpha+2\beta<2,\quad \alpha<1,
\]
the sampling drift dominates the transport, and, therefore,
\be
q={\beta\over\alpha+2\beta-1}
\label{4.13}
\ee
by (\ref{exit}) (and the discussion afterward).
Moreover,
since the sampling drift is asymptotically uniform in
space, the
displacement can be approximated asymptotically by
\be
\label{frozen_path}
\bmx^\ep(0)+\ep\int^{t/\ep^{2q}}_0 \vV(\bmx^\ep(0),s)ds.
\ee
Eq. (\ref{frozen_path}) is called
the space-freezing approximation,
in which the space dependence of
the Lagrangian velocity is suppressed.
Eq. (\ref{frozen_path}) defines a Gaussian process with
stationary increments.
It is easy to check that (\ref{frozen_path}) converges
to a fractional Brownian motion $\bB_H(t)$ by computing its
covariance tensor
\[
\langle\bB_H(t)\otimes\bB_H(t)\rangle={\bf C}
t^{2H}
\]
with the Hurst exponent
\[
H={1\over 2}+{\alpha+\beta-1\over 2\beta}=1/(2q)>{1\over 2}
\]
and the coefficient
\[
{\bf C}=E_0\int_{R^d}
{e^{-a_0|\bk|^{2\beta}}-1+a_0|\bk|^{2\beta}\over |\bk|^{2\alpha+4\beta-1}}
\left(\bI-\bk\otimes\bk|\bk|^{-2}\right)|\bk|^{1-d}d\bk.
\]

For a different and rigorous approach to the fractional Brownian motion limit,
see Reference \cite{fbm}.

\subsection{Regime V: subcritical cutoff}

If the cutoff is supercritical, $\delta \ll  k_c$, 
the sampling drift is effectively intact, so the frozen
path approximation
(\ref{frozen_path}) holds along with the
FBM limit with $q$ given by (\ref{4.13}).
 
If the cutoff is {\em subcritical}, $\delta\gg  k_c$,
the sampling drift
is absent and the transport is determined by
the fluctuating velocity field. 
We further decompose the subcritical wave numbers into 
those $|\bk|\sim \delta=\ep^\gamma$ and those much larger.
We have made the estimate ((\ref{upper2'}) for the contribution
of the latter to the eddy diffusivity. 

By a simple spectral calculation, the velocity field $\vU_\ep(\bx,t)$
consisting of the wave numbers $|\bk|\sim\delta$ can be
approximated by
\be
\label{subcut}
\delta^{1-\alpha}\vU(\delta \bx, \delta^{2\beta}t)
\ee
where $\vU$ has the energy spectrum (\ref{2.41}) for
$|\bk|\in [1,C]$ with a sufficiently large constant $C$.
Substituting (\ref{subcut}) into the equation of motion
we have
\be
\label{100}
d\bmx^\ep=
\ep^{\gamma(1-\alpha)-2q+1}
\vU(\bmx^\ep(t)/\vep^{1-\gamma},
t/\vep^{2q-2\beta\gamma})dt.
\ee
The effect of the wave numbers $|\bk|\gg\delta$ is like
adding a turbulent diffusivity 
to eq. (\ref{100}), i.e.
\be
\label{100''}
d\bmx^\ep=
\ep^{\gamma(1-\alpha)-2q+1}
\vU(\bmx^\ep(t)/\vep^{1-\gamma},
t/\vep^{2q-2\beta\gamma})dt+
\ep^{1-q}\sqrt{2\dot{\bD}^{(\ep)}}d\bB(t)
\ee
where $\bB(t)$ is the standard Browian motion
and $\dot{\bD}^{(\ep)}$ 
is the portion of the eddy diffusivity coming from the wave numbers
$|\bk|\gg\delta$ (cf. the discussion preceding
(\ref{upper2'})).

Since $\vU$ is a mixing flow, we expect 
the limit of (\ref{100''}) to be a Brownian motion.
We also expect the time variable in (\ref{100}) to dominate, so
we equate
$\ep^{\gamma(1-\alpha)-2q+1}=\vep^{2q-2\beta\gamma}$ and arrive at
the expression
\be
\label{q}
q=1+\gamma-\gamma(\alpha+\beta)
\ee
as one would expect from  a generalized
`diffusive' scaling of (\ref{100}).
We check that the space variable in (\ref{100}) is indeed relatively slow
in the sense
\be
\label{101}
(1-\gamma)/(q-\beta\gamma)<1,
\ee
for $\alpha+2\beta<2$.
With (\ref{q}), eq. (\ref{100}) becomes
\be
\label{100'}
d\bmx^\ep
=\eta_\veps^{-1}
\vU(\bmx/\eta_\ep^{(1-\gamma)/(q-\beta\gamma)}, t/\eta_\ep^{2})dt
+\ep^{1-q}\sqrt{2\dot{\bD}^{(\ep)}}d\bB(t),\quad
\hbox{with}\,\,\eta_\vep=\veps^{q-\beta\gamma}.
\ee
The bound (\ref{upper2'}) and (\ref{q}) imply that
$\ep^{1-q}\sqrt{2\dot{\bD}^{(\ep)}}\ll
\ep^{1-q+\gamma-\gamma(\alpha+\beta)}=1$ as $\ep$ tends to zero.
Subcriticality, $
\gamma< 1/(\alpha+2\beta-1),
$
implies
$\eta_\veps\to 0$.
Eq. (\ref{100'}) satisfies the conditions of
the diffusion limit theorem of \cite{Ko} and \cite{Ko2} for mixing
flows with a generalized `diffusive' scaling (\ref{100'})-(\ref{101}).
The limit is a Brownian motion with
the diffusion coefficients given by the Eulerian Taylor-Kubo formula 
\be
\label{TK2}
\int^\infty_0 \langle \vU(0,t)\otimes \vU(0,0)\rangle dt.
\ee
The condition $\alpha+\beta>1$ implies 
$q<1$.


The scaling law with $H=1/2$, $q$, given by
(\ref{q}), and the eddy diffusivity given by the Eulerian Taylor-Kubo
formula (\ref{TK2})
holds in the other part of Regime V ($\alpha\geq 1$) as well
(see Section 5).

\subsection{Regime II': critical cutoff}
When the infrared cutoff is critical i.e. $\gamma=\gamma_c=
(\alpha+2\beta-1)^{-1}$ the critical wave numbers are still present.
Therefore the scaling exponent is given by (\ref{4.13}).

Rescaling the velocity field
$\vU_\ep$  consisting
entirely of wave numbers $|\bk|\sim\ep^{\gamma_c}$ but $|\bk|< \ep^{\gamma_c}$
by (\ref{subcut}),
we have, instead of (\ref{100''}), the equation
\[
d\bmx^\ep=
\vU(\bmx^\ep(t)/\vep^{1-\gamma_c},
t)dt+
\ep^{1-q}\sqrt{2\bD^{(\ep)}}d\bB(t)
\]
where $\vU$ has the energy spectrum (\ref{2.14})
supported in $[1,\infty)$
and $\bD^{(\ep)}$ is the eddy diffusivity.

Since $\gamma_c>1$ and $q<1$, we have in the limit $\ep\to 0$
\be
\label{limit}
d\bZ(t)=\vU(0,t)dt.
\ee
Because the energy spectrum of $\vU$ does not have small wave numbers,
the process $\bZ$ is probably not self-similar.
Thus, the Hurst exponent is not well-defined. However,
the long time asymptotics of (\ref{limit}) is a Brownian motion,
due to the mixing property of $\vU$, so we may associate the
asymptotic Hurst exponent $1/2$ to
the process defined by (\ref{limit}).

\subsection{Regime III: time-freezing property}


In this regime, the 
sampling drift 
is a time dependent, mean zero
random variable whose second moment is of order
\[
\int^{k_c}_0|\bk|^{1-2\alpha}d|\bk|\sim \epsilon^{2-2\alpha}.
\]
The mixing time for the 
sampling drift is no less than
$1/\epsilon^{2\beta}$. On this time scale the
sampling drift transports
particles over the distances no less than
$\epsilon^{1-\alpha}\epsilon^{-2\beta}=\epsilon^{1-\alpha
-2\beta}
$
which is no less than the spatial observation scale $1/\epsilon$ if
$\alpha+2\beta\geq 2.$
Thus, the 
sampling drift appears steady on the observation scale.
The time to exit a ball of radius $1/\ep$, based on the  
sampling drift
alone is
$\epsilon^{-1}\epsilon^{\alpha-1}=\epsilon^{\alpha-2}.
$

From (\ref{up}) it follows
that the time to exit a ball of radius $1/\veps$, with its center moving 
by the sampling drift,
is 
$\ll\epsilon^{\alpha-2}$,
as a result of the spatially fluctuating,
subcritical wave numbers ($|\bk|\gg \ep$).
Thus, by (\ref{exit}), the effect of the subcritical wave numbers
is dominated by
that of the 
sampling drift.
Therefore, by (\ref{exit}), we have
\be
\label{3.5}
q=1-\alpha/2.
\ee
The scaling exponent (\ref{3.5}) is superdiffusive for $\alpha>0$.
Note that
\[
q-\beta/(\alpha+2\beta-1)=(\alpha/2+\beta-1)(1-\alpha)/(\alpha+2\beta-1)
<0
\]
for $1<\alpha+2\beta<2, \alpha<1$ and, thus, for the same $\alpha$,
the rate of transport
in Regime II is smaller (since faster
decorrelation in time tends to slow down the transport).

Since the transport is dominated by the effectively steady
sampling drift, we may consider the velocity field $\vU_\ep$ consisting 
entirely of the wave numbers $|\bk|\sim k_c=\ep$ and
freeze the time variable in the resulting velocity field as $\ep\to 0$.
Rescaling as in (\ref{subcut}), we have the equation
\be
\label{frozen2}
d\bmx^\ep(t)=\ep^{2-\alpha-2q}\vU(\bmx^\ep(t),0)dt
+\veps^{1-q}\sqrt{2\kappa}d\bmw(t).
\ee
Eq. (\ref{frozen2}) has a limit if and only if
(\ref{3.5}) holds,
which also make the diffusion term vanish
in the limit for $\alpha>0$. Formally, the limit process $\bZ$ satisfies
\be
\label{ultra}
d\bZ(t)=\vU(\bZ(t),0)dt
\ee
where $\vU$ has the energy spectrum (\ref{2.41}) supported 
in $|\bk|\in (0,\infty)$. The supercritical cutoff is now removed
by the rescaling (\ref{subcut}). This indicates the limit $\bZ$ is
self-similar.

It should be noted that
the velocity field $\vU$ is a generalized function
because of the ultraviolet divergence in the energy spectrum of $\vU$.
Thus, eq. (\ref{ultra}) is not well-defined in the ordinary sense.
Study of transport in generalized velocity fields is interesting
by itself, but we do not pursue it here. Our only purpose is to use 
the energy spectrum  of the velocity field 
as an indicator of the self-similarity of the limit process and to show
how the space and the time dependence of the velocity field
enter the equation.
The same remark applies to the same situation in the sequel as well as
(\ref{limit})
and will not be repeated.

We now identify the Hurst exponent of $\bZ$.
From (\ref{H'}), we have the asymptotics for
the covariance of the successive increments
on the time scale $t\sim\veps^{-2q}$
\be
\label{56}
\langle\left(\bmx(2t)-\bmx(t)\right)
\cdot\left(\bmx(t)-\bmx(0)\right)\rangle
\sim t^{2H}\sim\veps^{-4qH}.
\ee
On the other hand, by (\ref{exit}),
the covariance of the successive increments on the
time scale $t\sim\veps^{-2q}$ is of order
$(k_c^{1-\alpha}\veps^{-2q})^2.$
Equating it with (\ref{56}), we have $H=1/(2q)$. 
We hypothesize that the limit be a fractional
Brownian motion. 

Molecular diffusion probably has no significant effect on
the scaling law, even though the presence of molecular diffusion
is needed in the variational principle (\ref{eff}).
As remarked after (\ref{16''}) in Section~2, the molecular
diffusion is negligible for $\beta\leq 1$. For $\beta\geq 1$,
as larger $\beta$ gives rise to longer
correlation times,  the scaling law should have equal or smaller
scaling exponent $q$.
However, since the time-freezing property has already set in for $\beta<1$
and resulted in a scaling exponent independent of $\beta$, the absence
of molecular diffusion would not have changed the scaling exponent
for $\beta\geq 1$.

\subsection{Regime VI: subcritical cutoff}
The results of the previous section
hold for any supercritical cutoff
($\gamma>\gamma_c=1$) as the sampling drift is essentially intact and dominates
the transport.

We separate the wave numbers $|\bk|\sim \delta$ from $|\bk|\gg\delta$
and rescale the equation as in Section 4.3 to obtain (\ref{100''}).
As before, we expect the limit to be a Brownian motion. In this case,
however,  we expect the space variable in the velocity field to dominate the
transport. So we equate $\ep^{\gamma(1-\alpha)-2q+1}=\ep^{\gamma-1}$ and
obtain the scaling exponent
\be
\label{53'}
q=1-\gamma\alpha/2.
\ee
Rewriting (\ref{100''}) with $\eta_\ep=\ep^{1-\gamma}$ we have
\be
d\bmx^\vep(t)
=\eta_\veps^{-1}
\vU(\bmx^\ep(t)/\eta_\vep,
t/\eta_\veps^{2(q-\beta\gamma)/(1-\gamma)})dt
+
\ep^{1-q}\sqrt{2\dot{\bD}^{(\ep)}}d\bB(t)
+\vep^{1-q}
\sqrt{2\kappa}d\bmw(t).
\label{5.222}
\ee
The bound (\ref{upper2'}) and
(\ref{53'}) imply that $\ep^{1-q}\sqrt{2\dot{\bD}^{(\ep)}}
\ll \ep^{1-q-\gamma\alpha/2}=1$. Since $q<1$ the factor in front of
the molecular diffusion also tends to zero as $\ep\to 0$.
The time variable is relatively slow in  the sense
\[
\eta_\veps^{2(q-\beta\gamma)/(1-\gamma)}\gg \eta_\ep^2
\]
as $\alpha+2\beta>2$.

The velocity field $\vU$ has
a fixed infrared cutoff and, consequently, gives rise to a 
{\em space-homogeneous} vector potential, 
so, as suggested by the diffusion limit theorem of \cite{moser2},
the limit should be a Brownian motion.

The diffusion limit theorem of \cite{moser2}, however,
does not apply directly because
it was proved with a non-vanishing molecular diffusion.
Generalizing the theorem of \cite{moser2} to the situation 
with a vanishing molecular diffusion such as eq. (\ref{5.222})
remains a challenging
problem in turbulent transport.


 
\commentout{
The time-freezing approximation fails if $\gamma<q/\beta$; in
this case, the scaling exponent $q$ may be larger  than that
given by (\ref{53'}), because of decorrelation in time, and the
limit should also be a Brownian motion.
A borderline cutoff $\gamma=1$ is sometime made in the literature
(see, e.g., \cite{AM}, \cite{P}), and the corresponding results may
also be different.
}

\subsection{Regime III': critical cutoff}
When the cutoff is critical, i.e. $\gamma=1$, the critical wave numbers
are still present and dominates the transport. Therefore, the scaling
exponent is given by (\ref{3.5}).  

The difference is that eq. (\ref{frozen2}) now has the limit satisfying
eq. (\ref{ultra}) with
the energy spectrum of $\vU$ supported in
$|\bk|\in [1,\infty)$ rather than $(0,\infty)$. As a consequence,
the limit is not self-similar. 

\subsection{Phase boundary}
The transport for the phase boundary $\alpha+2\beta=2, 0<\alpha<1$
contrasts interestingly to regime on either side of the boundary.

When the cutoff is supercritical,  the  sampling drift is present
and dominates the transport.  Thus, we expect $q=\beta$ from  (\ref{4.13}).
Indeed, with that and $\gamma_c=1$, we have as before the asymptotic equation
\[
d\bmx^\ep=
\vU(\bmx^\ep(t),
t)dt+
\ep^{1-q}\sqrt{2\bD^{(\ep)}}d\bB(t)
\]
where  $\vU$ has the energy spectrum (\ref{2.14}) supported
in $|\bk|\in (0,\infty)$. In the limit, the diffusion term vanishes
(cf. (\ref{upper}))
and the limit $\bZ$ satisfies the equation
\be
\label{limit2}
d\bZ(t)=\vU(\bZ(t),t)dt
\ee
whose solution is expected to be a fractional Brownian motion.

When the cutoff is subcritical,  we expect $q=1-\gamma+\gamma\beta$
from (\ref{q}). With that and $\eta_\vep=\veps^{1-\gamma}$,
we have the asymptotic equation
\be
\label{equal}
d\bmx^\ep
=\eta_\veps^{-1}
\vU(\bmx/\eta_\ep, t/\eta_\ep^{2})dt
+\ep^{1-q}\sqrt{2\dot{\bD}^{(\ep)}}d\bB(t),
\ee
where the energy spectrum of $\vU$  is supported in $|\bk|\in [1,\infty)$.
The diffusion term  dies out in the limit as before
(cf. (\ref{upper2'}))
and the limit $\bZ$
is a Brownian motion due to the spectral gap in $\vU$ by a turbulent
diffusion theorem of \cite{markov}.

Contrary to Regimes II, II', V, III, III' and VI, both the time and the
space dependence of the velocity field in (\ref{limit2}) and
(\ref{equal}) affect the transport.

\section{Phase diagram for $\alpha>1$}
If the infrared cutoff threshold $\delta $ is fixed as
$\vep$ tends to zero then the velocity
is mixing in time and, by the turbulent diffusion theorem of \cite{markov},
the scaling limit
is a Brownian motion ($H=1/2$).

Anomalous scaling limits arise when $\delta$ is coupled to
the spatial observation scale:
$\delta=\vep^\gamma, \gamma>0.$
The exponent $\gamma$ characterizes the relation between the spatial
observation scale $1/\veps$ and the energy containing scale $1/\delta$.
And, in this case, superdiffusive scaling results from
divergent mean kinetic energy as the infrared cutoff is removed;
in particular, when, the cutoff is {\em supercritical},
$\gamma>\max{\{(\alpha+2\beta-1)^{-1}, 1\}}$,
(Regime IV), the energy containing scale
is larger than the spatial observation scale, and it results
in a super-ballistic scaling $q<1/2$ and a regular limit ($H=1$).

Since the transport is dominated by wave numbers $|\bk|\sim\delta$,
it is natural to
rescale the velocity $\vV$ 
as
\be
\label{5.1}
\vV(\bx,t)=\delta^{1-\alpha}\vU(\delta\bx,\delta^{2\beta}t)
\ee
where the velocity field $\vU$ has the energy spectrum (\ref{2.41})
supported in $|\bk|\in [1,\infty)$. Contrary to $\vU$ occurring
in the case of $\alpha\leq 1$, the velocity field $\vU$ in the case
of $\alpha>1$ is an ordinary function, since there is no ultraviolet or
infrared
divergence,
and temporally mixing due
to the spectral gap in $\vU$.

In terms of $\vU$,
the equation of motion becomes
\beq
d\bmx^\vep(t)&=&\delta^{1-\alpha}\vep^{1-2q}\vU(\delta
\bmx^\ep(t)/\veps,
\delta^{2\beta}t/\vep^{2q})dt\nonumber\\
&=&\ep^{\gamma(1-\alpha)-2q+1}
\vU(\bmx^\ep(t)/\vep^{1-\gamma},
t/\vep^{2q-2\beta\gamma})dt.
\label{5.2}
\eeq
Now that the infrared cutoff of $\vU$ is 1, the limit is expected to be a
Brownian motion as long as either space or time variable is
fast. Depending on the parameters,
two types of diffusion limit theorems in the literature are
pertinent to this limit: one is based on velocity decorrelation in time
(\cite{K},\cite{KP}, \cite{Ko}, \cite{Ko2}) and the other
on velocity decorrelation in space (\cite{moser2}).

Eliminating the infrared divergence by rescaling
is also the approach 
of Avellaneda and Majda \cite{AM}, in which the 
case of  a critical cutoff $\gamma=1$ was considered in the region
$\beta<1/2, 0<\alpha<2$. Here we adopt the same idea of rescaling
and generalize their results
by using new limit theorems
which were not available to them.

\subsection{$\gamma \geq 1$: Regimes V, IV and IV'}
For
$\gamma\geq 1$, eq. (\ref{5.2})  does not have fast space variables.
To have a nontrivial limit, we must have $2q-1+\gamma(\alpha-1)\geq 0$:
For
$\gamma\geq 1$ space variable is not fast in (\ref{5.2}).
To have a nontrivial limit, we must have $2q-1+\gamma(\alpha-1)> 0$
or $2q-1+\gamma(\alpha-1)=0$. The former case
gives rise to Regimes V wheres the latter gives rise to Regimes IV or
IV'.

\subsubsection{Regime V}
When 
\be
\label{5.5'}
2q-1+\gamma(\alpha-1)>0,
\ee,
$\vU$ in (\ref{5.2})
has a large multiplier, so a
nontrivial scaling limit requires rapid time  relaxation,
i.e.,  
\be
\label{5.5}
q>\gamma\beta. 
\ee
By choosing a generalized `diffusive' scaling for eq.
(\ref{5.2}), i.e., 
$(\delta^{\alpha-1}\vep^{2q-1})^{-1}=\delta^\beta/\vep^q$
or
\be
\label{5.4}
q=1+\gamma-\gamma(\alpha+\beta),
\ee
(\ref{5.2}) becomes
\be
d\bmx^\vep(t)
=\eta_\veps^{-1}
\vU(\eta_\veps^{(\gamma-1)/(q-\beta\gamma)}\bmx^\ep(t),
t/\eta_\vep^{2})dt
\label{57'}
\ee
with $\eta_\veps=\veps^{q-\beta\gamma}$.
Eq. (\ref{57'}) has the form of the classical 
diffusion limit theorem (\cite{K}, \cite{B}, \cite{KP})
(Moreover,
the velocity field $\vU$ is smooth
and satisfies the mixing condition of Rosenblatt (\cite{Ro})
even for $\beta>0$ since $\vU$ has no small $\bk$ components)
Thus,
the process $\bmx^\vep(t)$ converges 
to a Brownian motion ($H=1/2$)
with diffusion coefficients given by the Eulerian Taylor-Kubo
formula (\ref{TK2}).

However, there is one constraint to be considered:
(\ref{5.4}) must
be consistent with (\ref{5.5}), i.e., $\eta_\veps$ must
tend to zero with $\veps$. This means
$\gamma<1/(\alpha+2\beta-1)$, a subcritical cutoff.

\subsubsection{Regime IV: smooth motion}
If the cutoff is supercritical, 
as discussed in Section 3.1, the transport in this regime is dominated by
the sampling drift that is, in turn, dominated by the wave numbers
near by
the infrared cutoff.
Time as well as space dependence of the velocity field
are
irrelevant. 
Because both space and time variables are slow
in the velocity field,  nontrivial scaling limit  holds only  if
$\delta^{\alpha-1}\epsilon^{2q-1}=1$.
Thus we have
\be
\label{5.7}
q=(1+\gamma)/2-\gamma\alpha/2.
\ee
Consistency, $0<q<\gamma\beta$, then implies that
$\gamma>1/(\alpha+2\beta-1)$ and
\[
\alpha<1+1/\gamma.
\]
The limit process $\bZ(t)$ is
advected by a constant drift
\be
d\bZ(t)=\vU(0,0)dt
\label{106}
\ee
and is regular, or smooth ($H=1$).


We note that the limit
(\ref{106}) is independent of the initial condition $\bmx^\ep(0)$,
is self-similar and has a well defined Hurst exponent. 

\subsubsection{Regime IV': critical cutoff}
\label{iv'}
When
$2q-1+\gamma(\alpha-1)=0$,
or, equivalently,
\be
\label{7.5}
q=(1+\gamma)/2-\gamma\alpha/2,
\ee
and
\be
\label{7.6}
q=\gamma\beta
\ee
a nontrivial limit results.
Combining  (\ref{7.5}) and (\ref{7.6}), we have
$\alpha+2\beta=1+1/\gamma$ or
\be
\label{7.6'}
\gamma=1/(\alpha+2\beta-1)
\ee
which defines a critical cutoff for
$\alpha+2\beta <2$.
The limit $\bZ(t)$ 
satisfying
\be
\label{7.7}
d\bZ(t)=\vU(0,t)dt
\ee
is a smooth, Gaussian process. The Hurst exponent is not strictly
well-defined for $\bZ(t)$ due to lack of self-similarity.
On large time scales, however, (\ref{7.7}) has a Brownian motion limit
as $\vU$ is temporally mixing and, thus,
an asymptotic Hurst exponent $H=1/2$.
This case is similar to Regime V.

In particular, when $\gamma=1=1/(\alpha+2\beta-1)$ and $q=\beta$,
eq. (\ref{5.2})
is independent of $\ep$, i.e. $\bmx^\ep(t)=\bZ(t)$ with
\be
d\bZ(t)
=
\vU(\bZ(t),t)dt.
\label{ko}
\ee
The Hurst exponent is not strictly well-defined for $\bZ(t)$ due to
lack of self-similarity.
But, as the velocity field is temporally mixing,
the long-time limit of $\bZ(t)$ is a Brownian motion,
by the turbulent diffusion theorem of \cite{markov}.
So $H=1/2$ is the asymptotic Hurst exponent.

Another critical cutoff is $\gamma=1$ for $\alpha+2\beta> 2$.
The equation of motion (\ref{5.2}) becomes
\[
d\bmx^\vep(t)
=\ep^{2-\alpha-2q}
\vU(\bmx^\ep(t),
t/\vep^{2q-2\beta})dt
\]
which has a nontrivial limit when $q=\beta=1-\alpha/2$.
The limiting process $\bZ(t)$ satisfies
\be
d\bZ(t)=\vU(\bZ(t),0)dt
\label{106'}
\ee
which is regular for finite times, but is not self-similar.
Thus the
Hurst exponent is not well defined.
It is not clear whether an asymptotic Hurst exponent is well-defined 
either, since
we do not know if, without molecular diffusion, motion in three dimensional,
steady flows like (\ref{106'})
can be homogenized or not.
Bounded and unbounded streamlines may co-exist in steady flows and,
if so, the resulting limit would depend on initial conditions.


\subsection{$\gamma < 1$: Regimes V and VI}
With $\gamma<1$, fast space variables now enter the picture.
There are two regimes depending
on whether time relaxation dominates over space decorrelation.

\subsubsection{Regime V: dominant time relaxation}
This regime occurs when
$\delta^\beta/\vep^q\gg \delta/\vep
$
or, equivalently,
\be
\label{6.5}
q>1-\gamma+\gamma\beta.
\ee
Then, by choosing the `diffusive' scaling,
$(\delta^{\alpha-1}\vep^{2q-1})^{-1}=\delta^\beta/\vep^q,
$
i.e.,
\be
\label{6.1}
q=1+\gamma-\gamma(\alpha+\beta),
\ee
eq. (\ref{5.2})
can be rewritten as (\ref{57'}), except that the space
variable is also fast, albeit not fast enough to have an impact
in the diffusive scaling.
In this case, 
the generalized limit
theorems proved in
\cite{Ko},\cite{Ko2} are applicable and they extend the validity of
the Taylor-Kubo formula to our situation.
(Like the classical  diffusion limit theorem,
the generalized limit
theorems also require the mixing condition 
and regularity on the velocity $\vU$, both of which
are satisfied here.)
The limit is a Brownian motion ($H=1/2$) with
diffusion coefficients given by the Eulerian Taylor-Kubo
formula (\ref{TK2}).

Condition (\ref{6.5}) requires that
\be
\label{6.2}
\alpha+2\beta<2.
\ee


\subsubsection{Regime VI: dominant space decorrelation}
For
\be
\label{6.4}
q< 1-\gamma+\gamma\beta
\ee
velocity dependence on space now  dominates
over the dependence on time in the diffusive scaling of
eq. (\ref{5.2}).
By choosing the 
scaling 
$(\delta^{\alpha-1}\vep^{2q-1})^{-1}=\delta/\vep
$
or, equivalently,
\be
\label{6.3}
q=1-\gamma\alpha/2,
\ee
Eq. (\ref{5.2})
is rewritten as
\be
d\bmx^\vep(t)
=\eta_\veps^{-1}
\vU(\bmx^\ep(t)/\eta_\vep,
t/\eta_\veps^{2(q-\beta\gamma)/(1-\gamma)})dt,\quad\hbox{with}\,\,
\eta_\veps=\veps^{1-\gamma}.
\label{66}
\ee
The limit of (\ref{66}) should be 
a Brownian motion as $\vU$ gives rise to a space-homoegeous
vector potential
(cf. the discussion in 
Section~4.5).



Consistency ((\ref{6.4}) and $q>0$) requires that
$\alpha+2\beta>2,\quad \gamma<2/\alpha.$

\subsubsection{Phase boundary} 

On the phase boundary
$\alpha+2\beta=2$,
eq. (\ref{66}) becomes
\be
d\bmx^\vep(t)
=\eta_\veps^{-1}
\vU(\bmx^\ep(t)/\eta_\vep,
t/\eta_\veps^{2})dt
\label{66'}
\ee
with a temporally mixing flow $\vU$. 
Space and time correlations play comparable roles in 
(\ref{66'}). 
From the turbulent diffusion theorem for mixing flows (\cite{markov}),
it follows that the solution $\bmx^\ep(t)$
has a Brownian motion limit ($H=1/2$).

\section{Conclusions}
The supercritical and subcritical diagrams (Fig.1 and Fig.2)
are divided by the line,
$\alpha+\beta=1$,
and the line, $\alpha+2\beta=2$,
and/or the vertical lines
$\alpha=0, 1, 1+1/\gamma, 2/\gamma$.
First of all, $\alpha+\beta<1$ {\em or} $\alpha<0$ defines a cutoff independent
diffusive regime, in which the sampling drift is negligible.
Outside of the diffusive regime,
the line, $\alpha+\beta=1$, is the cross-over between short-ranged
and long-ranged velocity correlations; the latter manifests in the
fact that the sampling
drift dominates the transport and subcritical wave numbers are
negligible.
The line, $\alpha+2\beta=2$, is the cross-over
between velocity dependence on space and on time; in the region
above the line, velocity dependence
on time is negligible, whereas in the region below
the line velocity dependence on space is negligible.

Fig.3 and Fig.4 are cross sections of the full three-dimensional phase
diagram at $\gamma=const. >1$ and $\gamma=1$, respectively.
In Fig.3, the cutoff is
supercritical for $\alpha+2\beta> 1+1/\gamma$ and subcritical
for $\alpha+2\beta< 1+1/\gamma$. In Fig.4, the cutoff is subcritical
for $\alpha+2\beta<2$ and critical for $\alpha+2\beta\geq 2$.
Cutoffs with $\gamma<1$ are subcritical and, thus,
covered in Fig.2.

The limit is one of the three kinds: Brownian motion ($H=1/2$),
persistent 
fractional Brownian motion ($1/2<H<1$) or regular, or smooth, motion ($H=1$).
The relation $H=1/(2q)$
holds for $\alpha<1$ with supercritical infrared cutoff but neither
for subcritical cutoffs nor
for $\alpha>1$ (In these situations, $H<1/(2q)$, instead).
For the critical cutoff $\gamma=1$, the Hurst exponent is not well defined.
However,
an {\em asymptotic} Hurst exponent 
may be defined and it is equal to $1/2$.
The diffusive regime ($q=1, H=1/2$) is most robust in that
the scaling law is independent of any infrared cutoffs. The fractional
Brownian motion limit of Regime II and III
is not affected by supercritical cutoffs.
All other regimes are cutoff dependent. 

In the case of subcritical infrared cutoffs, with the rescaling of
the velocity field, the diagram can be understood by means of
three types of diffusion limit theorems in the
literature:
(i) one for which the spatial dependence of velocity is negligible
and the effective diffusivity is explicitly 
given by the Eulerian Taylor-Kubo formula 
(\cite{Kh},\cite{KP},\cite{Ko},\cite{Ko2}), (ii) another
for which the temporal dependence of velocity
is negligible, but molecular diffusion is assumed to be present,
and the effective diffusivity is implicitly given by 
a pair of variational principles (\cite{moser1}, \cite{moser2}, \cite{dt})
or a Stieltjes integral formula (\cite{AV}), and (iii) the other
for which space and time dependence of velocity play comparable roles
and is referred to as turbulent diffusion theorems (Two such theorems
are proved in \cite{markov}).
As in (ii), the turbulent eddy diffusivity can be written as a
variational principle similar to (\ref{2.22}).
After rescaling, Type (i) limit theorems apply to Regime V;
Type (ii) limit theorems apply to Regime VI and part of Regime I;
Type (iii) limit theorems apply to Regime I or on the
phase boundary $
\alpha+2\beta=2, 0<\alpha<2$.
All three types of limit theorems are insensitive to the dimension.
This may explain certain
similarity between the phase diagram for shear-layer flows of
\cite{AM0} or \cite{GZ} and the diagram
for isotropic flows with a {\em subcritical} cutoff.

Similarly, there should
be three types of fractional-Brownian-motion limit theorems:
one completely determined by Eulerian time decorrelation (Regime II),
another  determined by the space decorrelation (Regime III) and the other
determined by the dependence of velocity on both
space and time (The phase boundary $\alpha+2\beta=2, 0<\alpha<1$).

In the case of supercritical infrared cutoffs new phenomena emerge:
dominant sampling drift,
fractional
Brownian motion limits, critical infrared cutoff and related
cutoff dependent effects.
Although these phenomena are introduced and analyzed
for motion in three-dimensional, isotropic flows, they also arise
in two-dimensional flows or anisotropic flows such as
random shear-layer flows.
A difference lies in the role of molecular diffusion which is
much more prominent for anisotropic or two-dimensional flows
(see discussion in the next section).
New variational principles for the cutoff dependent
eddy diffusivity 
are formulated and used to obtain general bounds for the
eddy diffusivity.

Contrary to subcritical and supercritical cutoffs, regimes (II', III'
and IV') with
critical cutoffs produce limits that are not self-similar and do not
possess a well-defined Hurst exponent.

Scaling limits of turbulent transport in
flows with a {\em nonzero}
mean drift have different phase diagrams (see \cite{KB1},
\cite{Z}, \cite{AM2})
and will be reported in  a forthcoming paper.



\subsection{Role of molecular diffusion}

Molecular diffusion has at least two roles:
(i) to eliminate possible dependence of scaling limit on the initial
point, as particles may be trapped by closed or bounded streamlines,
so that the process may be homogenized;
(ii) to reduce dynamic velocity correlation in time and, thus,
change the scaling law to one with 
larger scaling exponent $q$.
The first role of molecular diffusion is prominent for transport
in steady flows; without it, localized streamlines would prevent
homogenization from  happening (see \cite{rced}). In this connection,
molecular diffusion also helps blend the effects
of streamlines of different scaling behaviors.
In the present work, we always assume homogenization and we focus on
the second role of molecular diffusion.

Molecular diffusion is negligible in Regimes II, III,
IV, where the sampling
drift dominates the transport as well as
Regimes V and part of I (i.e., $\alpha+\beta<1$ in the limit of
high Peclet number), 
where velocity decorrelation in time is
significant. But its effect is not so clear 
in Regimes VI 
and the other part of Regime I (i.e., $\alpha+\beta\geq 1, \alpha<0$), where
the spatially fluctuating wave numbers dominate.

In this regard, when high wave
numbers are negligible, as
for $\alpha +\beta> 1$, one can go further by
comparing the term representing molecular diffusion, $\kappa\Delta$, and
the term representing the flow, $\cA$, in (\ref{16''}),
and see that, for $\beta\leq 1$,  the effect of molecular
diffusion should not affect the scaling law.
As the velocity dependence on space dominates
over that on time, the scaling law is independent of
$\beta$. In view of the discussion in Section 4.4 on effect of molecular
diffusion, we expect the scaling law for $\beta>1$
can be extrapolated from that for $\beta\leq 1$ to conclude
the scaling law of Regime VI is independent of molecular diffusion
as well as $\beta$.

As $\alpha$ on the phase boundary ($\alpha=0$)
is
bigger than that 
in the region ($\alpha+\beta\geq 1, \alpha<0$), for given $\beta$,
the scaling exponent
$q$ in the region, with the absence
of molecular diffusion, should not be less than 
that of the phase boundary, 
which is $1$. Therefore, $q=1$ in this region even without molecular
diffusion.

\section{Appendix: variational principle for eddy diffusivity}
To derive the variational principle (\ref{2.22}) we 
 consider  a pair of period cell problems for an arbitrary constant unit
vector $\bfe$
\beq
\label{2.15}
\cA^{(\ep,n)}\chi^{(\ep,n)}_++(\bmc_\ep +\widetilde{\vV}^{(\ep,n)})\cdot\nabla\chi^{(\ep,n)}_++\widetilde{\vV}^{(\ep,n)}\cdot\bfe&=&0,
\quad\hbox{in}\,\,\Om^{(n)}
\\
\cA^{(\ep,n)}\chi^{(\ep,n)}_--(\bmc_\ep +\widetilde{\vV}^{(\ep,n)})\cdot\nabla\chi^{(\ep,n)}_--\widetilde{\vV}^{(\ep,n)}\cdot\bfe&=&0,
\quad\hbox{in}\,\,\Om^{(n)}
\label{2.16}
\eeq
where both $\chi^{(\ep,n)}_+$ and $\chi^{(\ep,n)}_-$ satisfy the periodic boundary condition.
Note that (\ref{2.16}) is simply the adjoint of (\ref{2.15})
as $\widetilde{\vV}^{(\ep,n)}$ is divergence free. 

Adding and subtracting  (\ref{2.15})
and (\ref{2.16})  we obtain
\beq
\label{2.17}
\cA^{(\ep,n)}\chi^{(\ep,n)}+(\bmc_\ep +\widetilde{\vV}^{(\ep,n)})
\cdot\nabla\chi^{(\ep,n)'}&=&0\\
\cA^{(\ep,n)}\chi^{(\ep,n)'}+(\bmc_\ep +\widetilde{\vV}^{(\ep,n)})\cdot\nabla\chi^{(\ep,n)}+
\widetilde{\vV}^{(\ep,n)}\cdot\bfe&=&0
\label{2.18}
\eeq
where
\be
\label{2.21}
\chi^{(\ep,n)}={1\over 2}(\chi^{(\ep,n)}_++\chi^{(\ep,n)}_-),\quad\chi^{(\ep,n)'}={1\over 2}(\chi^{(\ep,n)}_+-\chi^{(\ep,n)}_-).
\ee
First we establish some useful identities for 
\[
D^{(\ep,n)}(\bfe)=\bD^{(\ep,n)}\bfe\cdot\bfe=
-\langle \cA^{(\ep,n)}\chi^{(\ep,n)}_+\chi^{(\ep,n)}_+\rangle_n.
\]

\begin{prop}
\be
\label{2.20}
D^{(\ep,n)}(\bfe)=\langle\chi^{(\ep,n)}_+\widetilde{\vV}^{(\ep,n)}\cdot\bfe\rangle_n=
-\langle\chi^{(\ep,n)}_-\widetilde{\vV}^{(\ep,n)}\cdot\bfe\rangle_n=-\langle\cA^{(\ep,n)}\chi^{(\ep,n)}_-\chi^{(\ep,n)}_-\rangle_n.
\ee
\end{prop}
The first  identity in 
(\ref{2.20}) follows from integration by parts after multiplication
of eq. (\ref{2.15}) by $\chi^{(\ep,n)}_+$. To verify the
second, we make use eqs. (\ref{2.15}), (\ref{2.16}) and the divergence free property
of $\widetilde{\vV}^{(\ep,n)}$ in the following calculation
\beq
\nonumber
D^{(\ep,n)}(\bfe)&=&\langle(-\cA^{(\ep,n)}-(\bmc_\ep +\widetilde{\vV}^{(\ep,n)})\cdot\nabla)\chi^{(\ep,n)}_+\chi^{(\ep,n)}_+\rangle_n\\
\nonumber
&=&\langle[-\cA^{(\ep,n)}\chi^{(\ep,n)}_+-(\bmc_\ep +\widetilde{\vV}^{(\ep,n)})\cdot\nabla\chi^{(\ep,n)}_+-\widetilde{\vV}^{(\ep,n)}\cdot\bfe]\chi^{(\ep,n)}_+\rangle_n+
\langle\chi^{(\ep,n)}_+\widetilde{\vV}^{(\ep,n)}\cdot\bfe\rangle_n\\
\nonumber
&{(\ref{2.15})\atop =}&\langle[-\cA^{(\ep,n)}\chi^{(\ep,n)}_+-
(\bmc_\ep +\widetilde{\vV}^{(\ep,n)})\cdot\nabla\chi^{(\ep,n)}_+-\widetilde{\vV}^{(\ep,n)}
\cdot\bfe]\chi^{(\ep,n)}_-\rangle_n+
\langle\chi^{(\ep,n)}_+\widetilde{\vV}^{(\ep,n)}\cdot\bfe\rangle_n\\
\nonumber
&=&\langle[-\cA^{(\ep,n)}\chi^{(\ep,n)}_-+(\bmc_\ep +\widetilde{\vV}^{(\ep,n)})\cdot
\nabla\chi^{(\ep,n)}_-]\chi^{(\ep,n)}_+\rangle_n
-\langle\chi^{(\ep,n)}_-\widetilde{\vV}^{(\ep,n)}\cdot\bfe\rangle_n+
\langle\chi^{(\ep,n)}_+\widetilde{\vV}^{(\ep,n)}\cdot\bfe\rangle_n\\
\nonumber
&=&
\langle[-\cA^{(\ep,n)}\chi^{(\ep,n)}_-+(\bmc_\ep +\widetilde{\vV}^{(\ep,n)})
\cdot\nabla\chi^{(\ep,n)}_-+\widetilde{\vV}^{(\ep,n)}\cdot\bfe]\chi^{(\ep,n)}_+\rangle_n-
\langle\chi^{(\ep,n)}_-\widetilde{\vV}^{(\ep,n)}\cdot\bfe\rangle_n\\
\nonumber
&{(\ref{2.16})\atop =}&-\langle\chi^{(\ep,n)}_-\widetilde{\vV}^{(\ep,n)}\cdot\bfe\rangle_n.
\eeq
Here we have used the identity
\[
\langle\left[(\bmc_\ep +\widetilde{\vV}^{(\ep,n)})\cdot\nabla\chi^{(\ep,n)}_+\right]
\chi^{(\ep,n)}_+\rangle_n={1\over 2}
\nabla\cdot\langle(\bmc_\ep +\widetilde{\vV}^{(\ep,n)})
(\chi^{(\ep,n)}_+)^2\rangle_n =0
\]
as a result of the incompressibility of $\bmc_\ep +\widetilde{\vV}^{(\ep,n)}$
and  the space-homogeneity of  $
\langle(\bmc_\ep +\widetilde{\vV}^{(\ep,n)})
(\chi^{(\ep,n)}_+)^2\rangle_n$.

Thus, in view of (\ref{2.21}), the following result is clear.
\begin{prop}
\[
D^{(\ep,n)}(\bfe)=\langle\chi^{(\ep,n)'}\widetilde{\vV}^{(\ep,n)}\cdot\bfe\rangle_n=-\langle\cA^{(\ep,n)}\chi^{(\ep,n)'}\chi^{(\ep,n)'}\rangle_n-\langle\widetilde{\vV}^{(\ep,n)}\cdot\nabla\chi^{(\ep,n)}\chi^{(\ep,n)'}\rangle_n.
\]
\end{prop}

Next we derive the variational principle (\ref{2.22}).

Let $g$ be the minimizer  of the convex functional in
(\ref{2.22}) and $g'$ be the periodic solution of the equation
\be
\label{2.25'}
\cA^{(\ep,n)} g'+(\bmc_\ep +\widetilde{\vV}^{(\ep,n)})\cdot\nabla g+
\widetilde{\vV}^{(\ep,n)}\cdot\bfe=0.
\ee
Taking the first variation of the functional in (\ref{2.22}) at $g$ we have
\be
\label{2.23}
-\langle\cA^{(\ep,n)} g\delta g\rangle_n-\langle\cA^{(\ep,n)} g'\delta g'\rangle_n=0
\ee
where the variation $\delta g'$ is related to the variation $\delta g$ by
\be
\label{2.24}
\cA^{(\ep,n)} \delta g'+(\bmc_\ep +\widetilde{\vV}^{(\ep,n)})\cdot\nabla\delta g=0
\ee
following (\ref{2.25'}).
Substituting (\ref{2.24}) into (\ref{2.23}) and integrating by parts we get
\[
\langle\cA^{(\ep,n)} g\delta g\rangle_n+\langle
(\bmc_\ep +\widetilde{\vV}^{(\ep,n)})\cdot\nabla g'\delta g\rangle_n=0
\]
for all admissible variations $\delta g$. Thus
\be
\label{2.26}
\cA^{(\ep,n)} g+(\bmc_\ep +\widetilde{\vV}^{(\ep,n)})\cdot\nabla g'=0.
\ee
Since eqs. (\ref{2.17}) and (\ref{2.18}) (also (\ref{2.25'}), (\ref{2.26}))
 are well posed,
we conclude
that $g=\chi^{(\ep,n)},g'=\chi^{(\ep,n)'}$ up to constants. 

By reversing
the above argument,  it is easy to see that
$g=\chi^{(\ep,n)}, g'=\chi^{(\ep,n)'}$ with $\chi^{(\ep,n)}, \chi^{(\ep,n)'}$ given by
(\ref{2.21}) are the minimizer of (\ref{2.22}).

\bigskip
{\bf Acknowledgement}
I thank an anonymous referee for careful reading
and useful remarks which led to
improvement of the manuscript.
The research is supported in part by National Science Foundation
Grant DMS-9600119.

\newpage

\centerline{\large\bf Guide to the phase diagrams}
\vspace{.5cm}

The full diagram is in the three dimensional space of $\alpha,\beta,\gamma$.
To simplify the presentation, we choose to portray the full diagram
as four planar diagrams: the supercritical diagram,
the subcritical diagram, the cross sections $\gamma={\em const.}>1$
and $\gamma=1$. Note that Regime II' does not show up
in any of the diagrams.
\begin{itemize}
\item The exponents $\alpha$ and $\beta$ defined in (\ref{1.1})-(\ref{2.41})
characterize the space-time correlations.
\item The exponent $\gamma$
is related to the infrared cutoff $\delta$ in (\ref{1.3})
as $\delta=\ep^\gamma$.
\item The critical exponent $\gamma_c=\max{\{1, (\alpha+2\beta-1)^{-1}\}}$.  
\item The scaling exponent $q$ and the Hurst exponent $H$ are defined
by (\ref{2.2}) and (\ref{H'}), respectively.
\item 
\begin{description}
\item[Regime I:] $\alpha+\beta< 1$ or $\alpha<0$.
\item[Regime II:] $\alpha+\beta>1,\alpha+2\beta<2,\alpha<1$ with
$\gamma>1/(\alpha+\beta-1)$.
\item[Regime II':] $\alpha+\beta>1, \alpha+2\beta<2, \alpha<1$ with
$\gamma=(\alpha+2\beta-1)^{-1}$.
\item[Regime III:] $\alpha+2\beta>2, 0<\alpha<1$ with $\gamma>1$.
\item[Regime III':] $\alpha+2\beta>2, 0<\alpha<1$ with $\gamma=1$.
\item[Regime IV:] $1<\alpha<1+1/\gamma$
with $\gamma> \max{\{1, (\alpha+\beta-1)^{-1}\}}$.
\item[Regime IV':] $1<\alpha<1+1/\gamma$ with $\gamma=\gamma_c$. 
\item[Regime V:] $\alpha+\beta>1,\alpha+2\beta<2$ with 
$\gamma<1/(\alpha+\beta-1)$.
\item[Regime VI:] $\alpha+2\beta> 2,\alpha\geq 0$ with $\gamma<1$.
\end{description}
\end{itemize}

\newpage

\vspace{5mm}

\begin{figure}
\begin{picture}(400,250)
\includegraphics{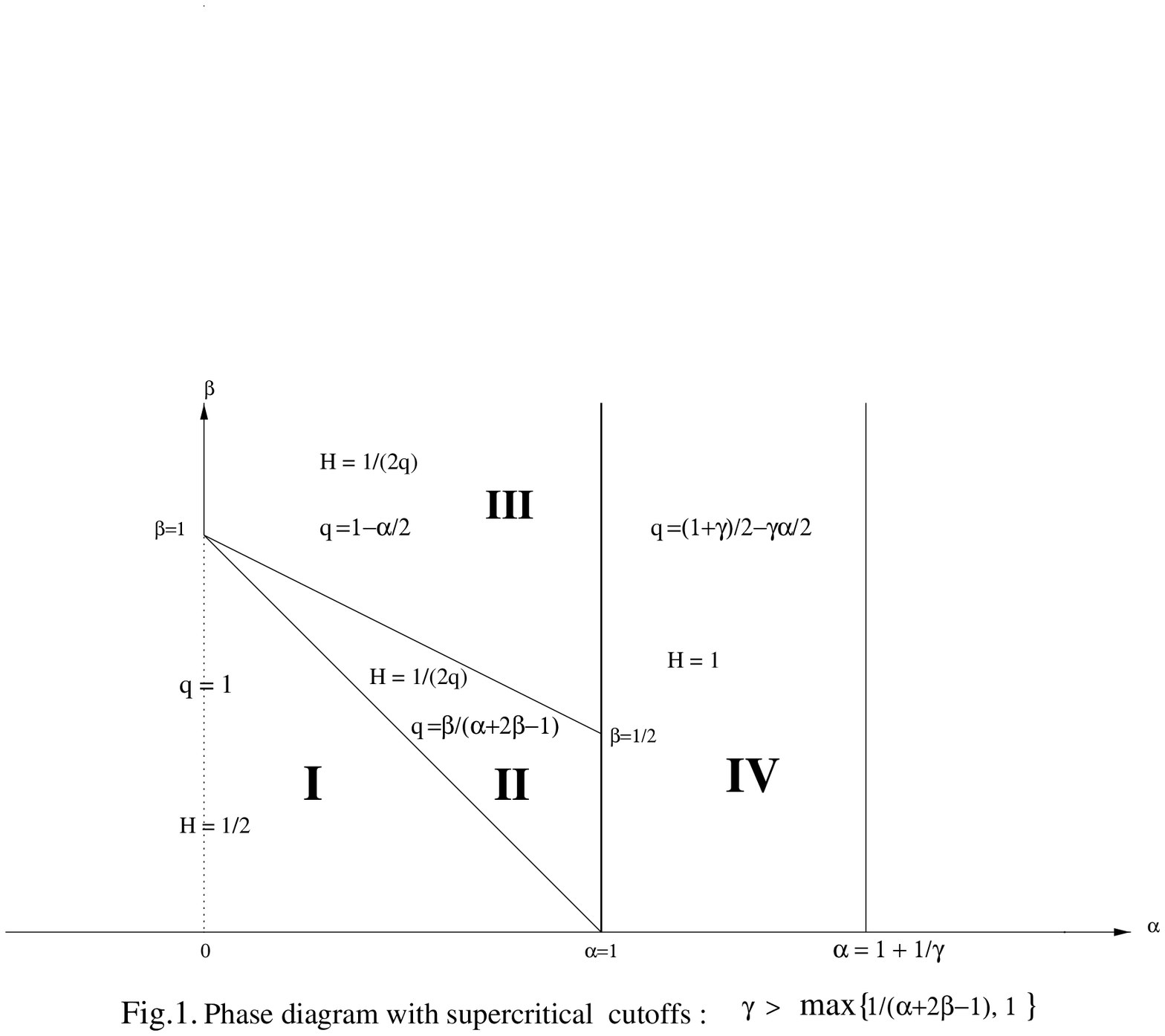}
\end{picture}
\end{figure}

\begin{figure}
\begin{picture}(400,250)
\includegraphics{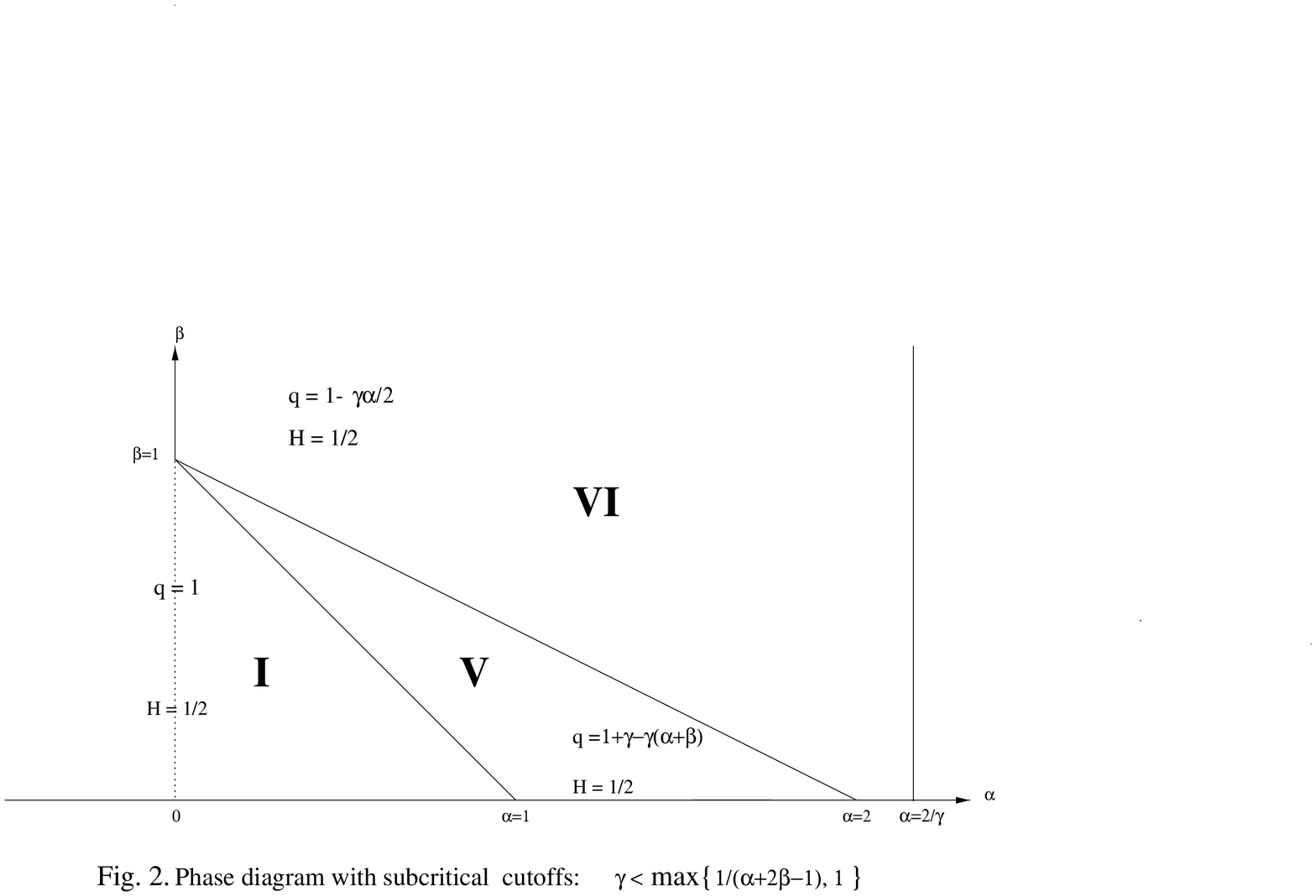}
\end{picture}
\end{figure}

\newpage

\vspace{5mm}

\begin{figure}
\begin{picture}(400,250)
\includegraphics{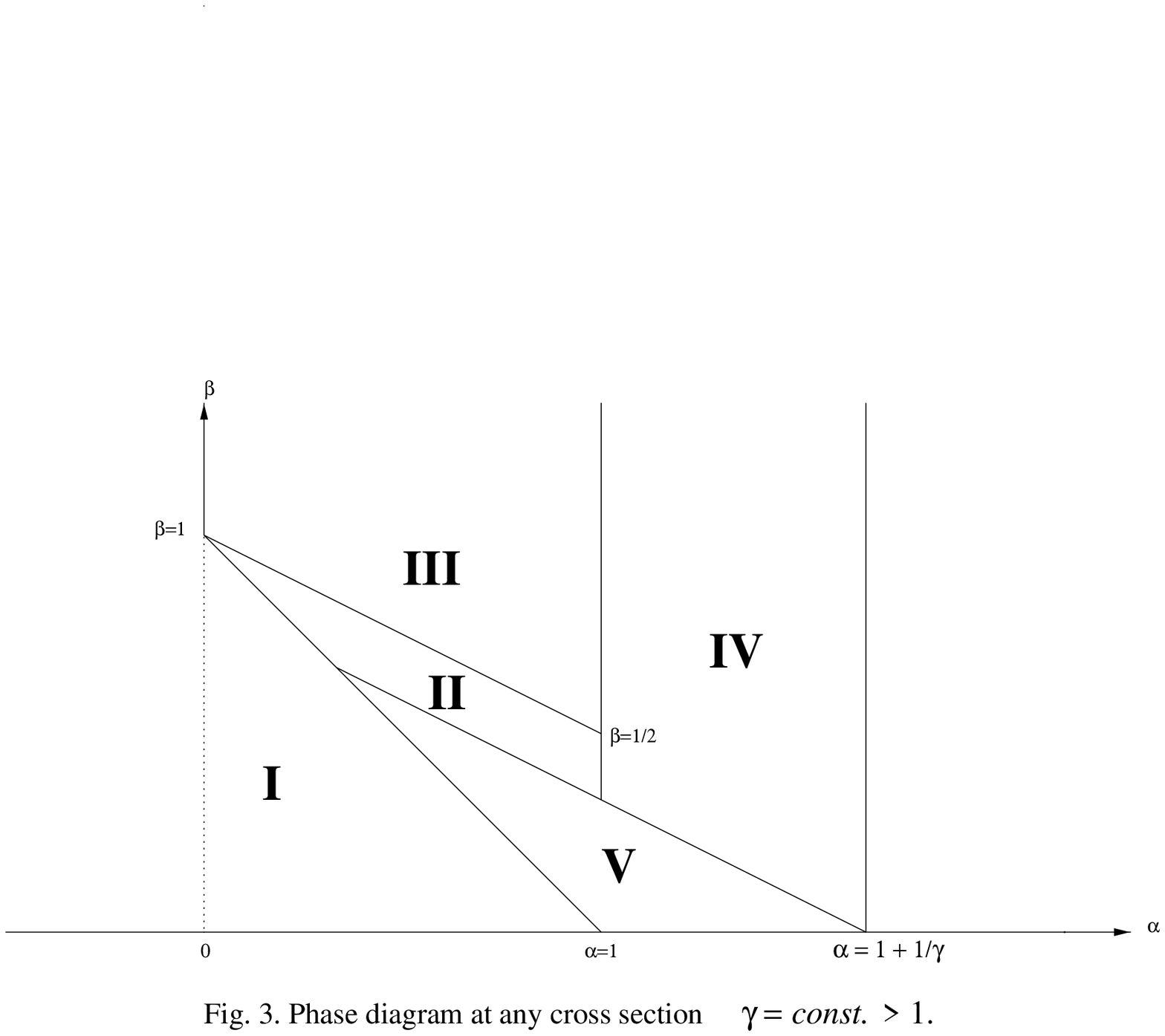}
\end{picture}
\end{figure}

\begin{figure}
\begin{picture}(400,250)
\includegraphics{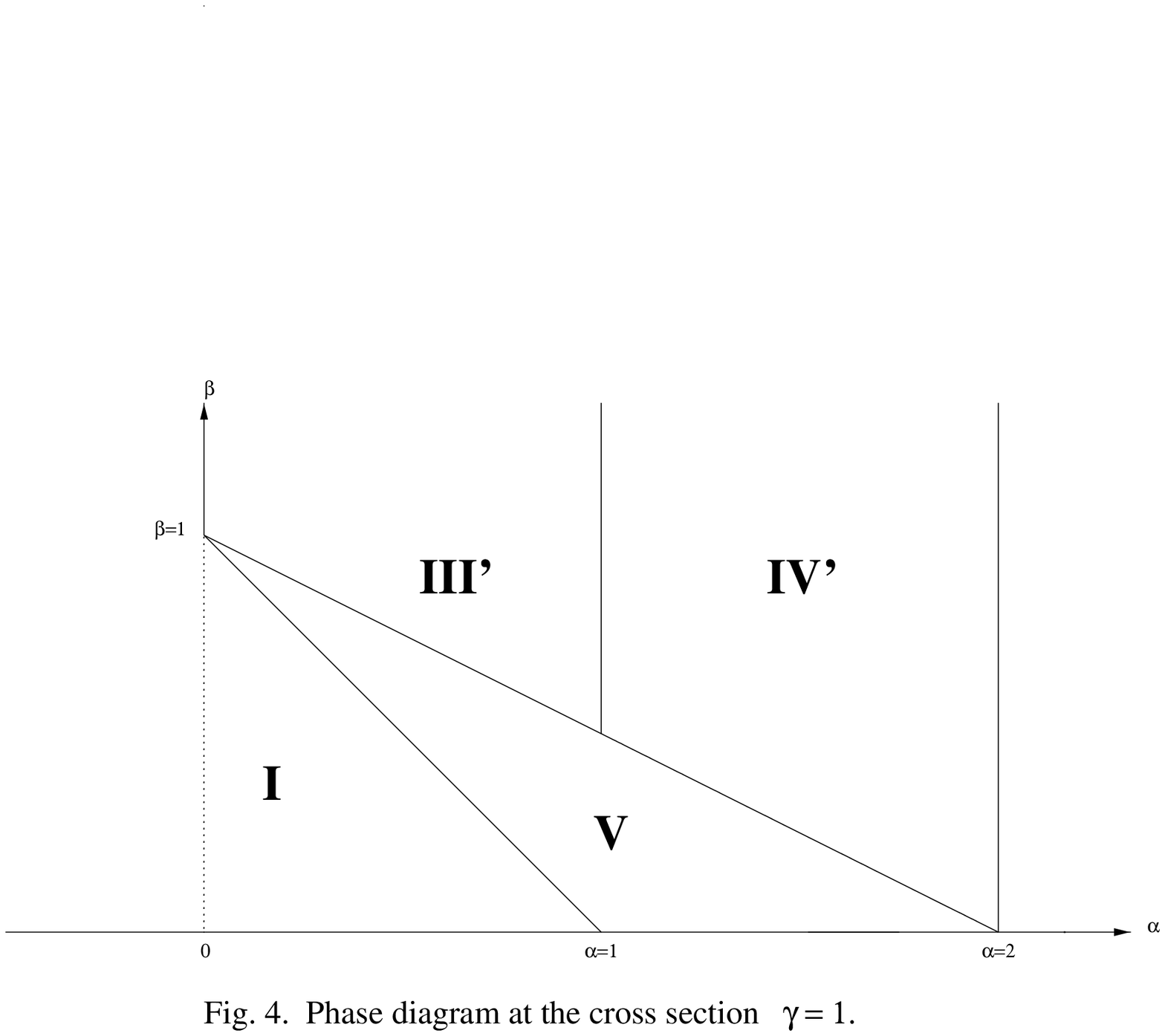}
\end{picture}
\end{figure}

\end{document}